\documentclass[manuscript]{emulateapj}
\pdfoutput=1
\usepackage{natbib}
\usepackage{graphicx}
\usepackage{xspace}

\newcommand{\um}{\,$\mu$m}

\newcommand{\hubble}{$H_0=72$ km s$^{-1}$ Mpc$^{-1}$}
\newcommand{\sunrise}{\textsc{sunrise}\xspace}
\newcommand{\gadgetthree}{\textsc{gadget-3}\xspace}

\shorttitle{SIGS-Simulation Paper I} 
\shortauthors{Lanz et al.}

\begin{document}

\title{Simulated Galaxy Interactions as Probes of Merger Spectral Energy Distributions}
\author{Lauranne Lanz\altaffilmark{1,2},  Christopher C. Hayward\altaffilmark{3},
Andreas Zezas\altaffilmark{1,4,5},  Howard A. Smith\altaffilmark{1}, 
Matthew L. N. Ashby\altaffilmark{1}, Nicola Brassington\altaffilmark{6}, Giovanni G. Fazio\altaffilmark{1},
 Lars Hernquist\altaffilmark{1}}

\altaffiltext{1}{Harvard-Smithsonian Center for Astrophysics, 60 Garden St.,  Cambridge, MA 02138, USA}
\altaffiltext{2}{Infrared Processing and Analysis Center, California Institute of Technology, MC100-22, Pasadena, California 91125, USA; llanz@ipac.caltech.edu}
\altaffiltext{3}{Heidelberger Institut f\"{u}r Theoretische Studien, Schloss-Wolfsbrunnenweg 35, 69118 Heidelberg, Germany}
\altaffiltext{4}{University of Crete, Physics Department \& Institute of Theoretical \& Computational Physics, 71003 Heraklion, Crete, Greece}
\altaffiltext{5}{Foundation for Research and Technology-Hellas, 71110 Heraklion, Crete, Greece}
\altaffiltext{6}{School of Physics, Astronomy and Mathematics, University of Hertfordshire, College Lane, Hatfield, AL10 9AB, UK}

\begin{abstract}
We present the first systematic comparison of ultraviolet-millimeter spectral energy distributions (SEDs) of observed and simulated interacting galaxies. Our sample is drawn from the Spitzer Interacting Galaxy Survey, and probes a range of galaxy interaction parameters. We use 31 galaxies in 14 systems which have been observed with \emph{Herschel}, \emph{Spitzer}, \emph{GALEX}, and 2MASS. We create a suite of \gadgetthree hydrodynamic simulations of isolated and interacting galaxies with stellar masses comparable to those in our sample of interacting galaxies. Photometry for the simulated systems is then calculated with the \sunrise radiative transfer code for comparison with the observed systems.  For most of the observed systems, one or more of the simulated SEDs match reasonably well. The best matches recover the infrared luminosity and the star formation rate of the observed systems, and the more massive systems preferentially match SEDs from simulations of more massive galaxies. The most morphologically distorted systems in our sample are best matched to simulated SEDs close to coalescence, while less evolved systems match well with SEDs over a wide range of interaction stages, suggesting that an SED alone is insufficient to identify interaction stage except during the most active phases in strongly interacting systems. This result is supported by our finding that the SEDs calculated for simulated systems vary little over the interaction sequence.
\end{abstract}


\section{INTRODUCTION}

Galaxy interactions, especially major mergers, are responsible for some of the most dramatic activity seen in galaxies. In the canonical view, interactions stimulate star formation, thereby powering the high infrared (IR) luminosities often seen in such systems \citep[e.g.,][]{vei02}: driving gas inflows to the central regions, resulting in heightened activity of the central supermassive black hole and local starburst activity \citep[e.g.,][]{dim05, spr05b}, and leading to significant morphological distortions \citep[e.g.,][]{hop06, mih94, mih96}. These activities, however, occur over timescales that make detecting evolution in individual systems or tracing the corresponding development in physical processes impossible. Hydrodynamic simulations of interacting galaxies provide a means of probing the interaction sequence and bypassing the problem of the timescales. 

A crucial test of any simulation is its ability to reproduce observations. Hydrodynamic simulations of galaxy interactions have primarily been tested in two ways:  by how well they reproduce the (optical) morphological distortions seen in such systems, and by how closely their simulated emission matches that of real systems. Some simulations are designed to reproduce specific systems \citep[e.g.,][]{privon13, karl13}, while others compare specific properties, such as colors, of a suite of simulations to observations \citep[e.g.,][]{sny13, jon10}.

\citet{too72} were the first to systematically model the morphologies of interacting galaxies. They used simple simulations of massless particles around two masses  to reproduce the tails and bridges seen in systems like the M51, the Mice (NGC\,4676) and the Antennae (NGC\,4038/4039). Much more recently, \citet[see also Barnes 2011]{bar09} developed Identikit, a modeling tool that uses N-body simulations to reproduce the morphology and kinematics of tidal tails in interacting systems. \citet{privon13} demonstrated Identikit's ability to reproduce the morphology and H I kinematics of NGC\,5257/5258, the Mice, the Antennae, and NGC\,2623 and to estimate the time since the first pericenter passage and until coalescence. \nocite{bar11}

Morphological analyses like these inherently suffer from an obvious bias: simulations trace mass but observations trace light. Better comparisons propagate light from the simulated luminous matter to a fiducial observer. \sunrise  \citep{jon06} accomplishes exactly that. It is a radiative transfer code that propagates the emission of simulated stars and active galactic nuclei (AGN) through a dusty interstellar medium (ISM) generated by the hydrodynamic simulations. It is an ideal tool for creating simulated spectral energy distributions (SEDs) for comparison to photometry. For example, \citet{jon10} simulated the SEDs of seven isolated galaxies, which they compared to Spitzer Infrared Nearby Galaxies Survey \citep[SINGS;][]{ken03} galaxies from \citet{dale07}. The \citet{jon10} simulations did not cover all of the parameter space spanned by SINGS; nonetheless, good matches from the SINGS sample were found for each of the simulated galaxies, demonstrating the ability of \sunrise to produce realistic galaxy SEDs. \citet{karl13} combined analyses of the morphology and emission, by creating a set of hydrodynamic simulations to reproduce the morphology of the Antennae and performing radiative transfer to determine the predicted emission in the Herschel Space Observatory's Photodetector Array Camera and Spectrometer (PACS) bands. 

This paper is the first systematic comparison of the observed and simulated SEDs spanning the range from the ultraviolet (UV) to the far-IR (FIR) for interacting galaxies. \citet{jon10} determined which of the SINGS galaxies were best reproduced by the SEDs of their simulated isolated spiral galaxies.  Our study takes a related but different approach: we do not create simulations aimed at specifically reproducing our observed interactions; rather, we produce a range of simulated interactions and examine how realistically they reproduce observed systems, determine which of the  simulated SEDs best reproduce the observed SEDs of interacting systems, and identify the simulation properties, such as stellar mass, SFR, or interaction stage, common to the set. 

At high redshifts, morphological details become impossible to resolve and so estimates of interaction stage based on morphology  \citep[e.g.,][]{dopita02} become unworkable. A spectral marker for interaction stage would be a powerful tool for examining how  interactions at high redshift compare to local interactions. Therefore, we ask whether there is an unambiguous signature of the interaction stage in the SED. In this paper, we compare the SEDs of a suite of simulations of interacting and isolated spiral galaxies to the SEDs of  31 interacting galaxies to examine the simulation properties able to reproduce the SED of an observed system. A clear extension of our study is to test whether there is a signature of the morphology in the SED by finding common morphology either within the set of best matches or between the matches and the observation.

 This article is organized as follows. We summarize our sample selection and the photometry  in Section 2. In Section 3, we describe the hydrodynamic simulations and the radiative transfer done in post-processing.  We discuss our matching methodology and the best-matched SEDs in Section 4. Section 5 contains a  discussion of the origins of the best- and worst-matched SEDs,  a comparison between the stellar and dust masses,  dust luminosity, star formation rate (SFR), specific star formation rate (sSFR) of the observed systems and the best-matched simulated counterparts, an analysis of the effectiveness of morphology-based interaction stage classification scheme, and an examination of the evolution of SEDs in major mergers. We summarize our results in Section 6.

\begin{figure}[bhp]
\centerline{\includegraphics[width=\linewidth]{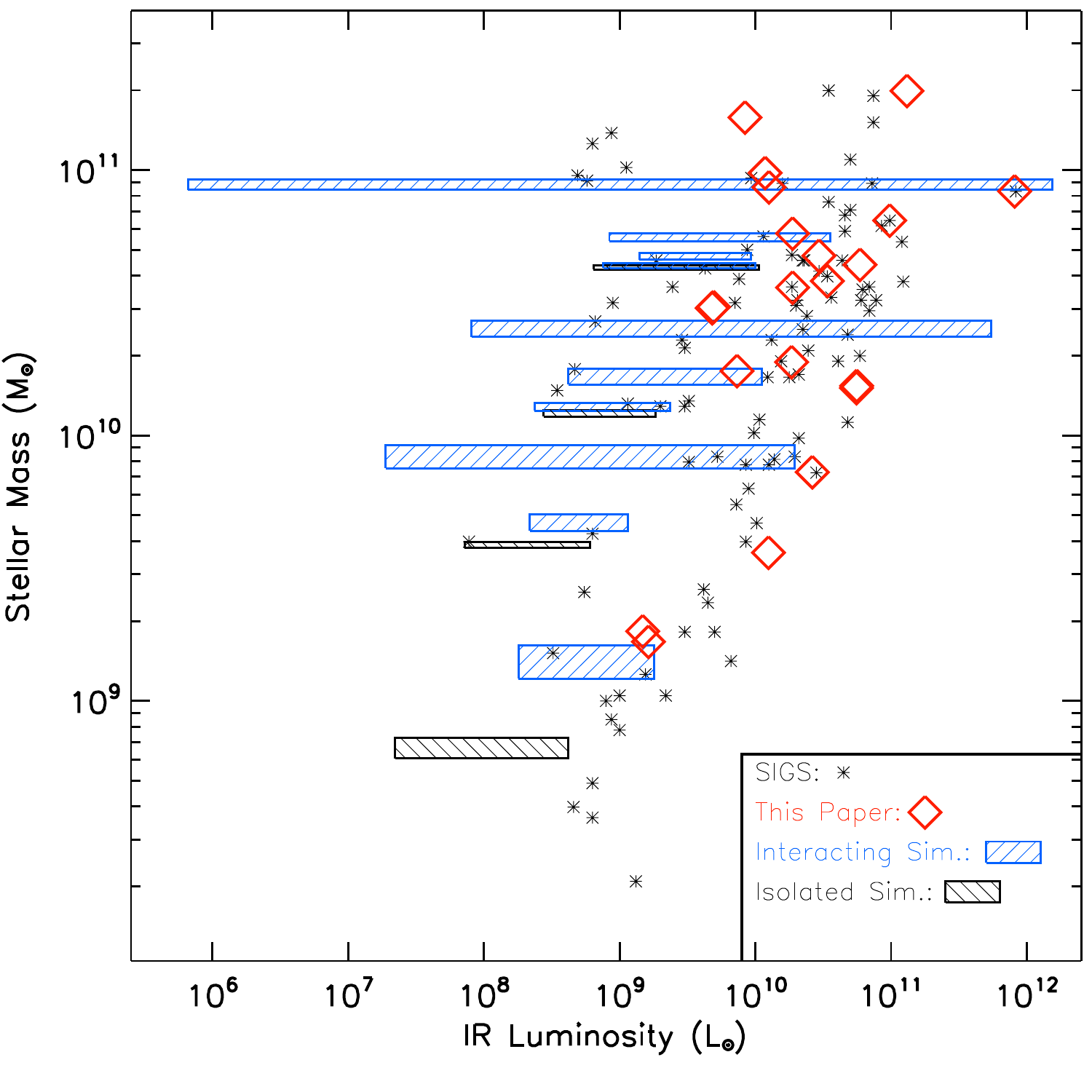}}
\caption{The range of IR luminosity and stellar mass covered by the observed sample used in this paper (red diamonds), the full parent SIGS sample (black stars), and the simulations (boxes in blue (interactions) and black (isolated galaxies)). The simulations cover the full range of the observed IR luminosity, and their mass range spans more than two orders of magnitude and are representative of most of the observed sample. The widths and height of a simulation box shows the range of the IR luminosity and stellar mass over the simulation. }
\label{par_range}
\end{figure}

\section{OBSERVATIONS}

Our sample and observations are described in detail in \citet[][Paper I]{lanz13}. Here we summarize the selection criteria for our galaxies and briefly describe the photometry and the fitting process that provides the stellar masses, dust masses and temperatures, SFR, and sSFR that we will compare to the simulations.

\begin{deluxetable*}{llrrccl}
\tabletypesize{\scriptsize}
\tablecaption{Sample Description\label{sample}}
\tablewidth{0.9\linewidth}
\tablecolumns{7}
\tablehead{
\colhead{} & \colhead{} & \colhead{R.A.} & \colhead{Decl.} 
& \colhead{Distance} & \colhead{Interaction} & \colhead{} \\
\colhead{Group} & \colhead{Galaxy} & \colhead{(J2000)} & \colhead{(J2000)} 
& \colhead{(Mpc)} & \colhead{Stage} & \colhead{Survey}  \\
\colhead{(1)} & \colhead{(2)} & \colhead{(3)} & \colhead{(4)} & \colhead{(5)} & \colhead{(6)}  & \colhead{(7)} 
}
\startdata
 1 	& NGC\,2976$^{+}$	     	& 	09 47 16.3	& +67 54 52.0	&	  3.75	& 2 	&  SINGS/KINGFISH \\  
  	& NGC\,3031		     	& 	09 55 33.2	& +69 03 57.9	&	  3.77	& 2 	&  SINGS/VNGS \\
	& NGC\,3034		     	& 	09 55 52.2	& +69 40 47.8   &	  3.89	& 2 	&  SINGS/VNGS \\
	& NGC\,3077$^{+}$	     	&	10 03 19.8	& +68 44 01.5	&	  3.93	& 2 	&  KINGFISH \\
2	& NGC\,3185		     	&	10 17 38.7	& +21 41 16.2	&	22.6~~~	& 2 	&  KINGFISH \\     
	& NGC\,3187		     	&	10 17 48.4	& +21 52 30.9	&	26.1~~~	& 3 	&  KINGFISH \\
	& NGC\,3190		     	& 	10 18 05.7	& +21 49 57.0	&	22.5~~~	& 3 	&  SINGS/KINGFISH \\   
3	& NGC\,3226		     	&	10 23 27.0	& +19 53 53.2	&	23.3~~~	& 4 	&  HRS \\
	& NGC\,3227		     	&	10 23 30.5	& +19 51 55.1	&	20.6~~~   	& 4 	&  HRS \\  
4	& NGC\,3395		 	&	10 49 50.0	& +32 58 55.2	&	27.7~~~	& 4 	&  HRS/SHINING \\ 
	& NGC\,3396  			&	10 49 55.2	& +32 59 25.7	&	27.7~~~	& 4 	&  HRS/SHINING \\ 
5	& NGC\,3424		     	&	10 51 46.9	& +32 54 04.1	&	26.1~~~	& 2 	&  HRS \\   
	& NGC\,3430		   	&	10 52 11.5	& +32 57 05.0	&	26.7~~~	& 2 	&  HRS \\ 
6	& NGC\,3448		     	&	10 54 38.7	& +54 18 21.0	&	24.4~~~	& 3 	&  HRS  \\ 
	& UGC\,6016$^{+}$	     	&	10 54 13.4	& +54 17 15.5	& $27.2^{*}$~~ & 3 	&  HRS \\ 
7	& NGC\,3690/IC\,694	&	11 28 31.2	& +58 33 46.7	& $48.1^{*}$~~ & 4 	&  SHINING\\ 
8 	& NGC\,3786		     	&	11 39 42.5	& +31 54 34.2	&	41.7~~~	& 3 	& \\
	& NGC\,3788		     	&	11 39 44.6	& +31 55 54.3  	&	36.5~~~	& 3 	& \\
9	& NGC\,4038/4039		&	12 01 53.9	& $-$18 52 34.8 &    	25.4~~~	& 4 	& VNGS/SHINING \\ 
10	& NGC\,4618$^{+}$	     	&	12 41 32.8	& +41 08 44.4	&	  7.28	& 3 	& KINGFISH \\ 
 	& NGC\,4625$^{+}$	     	&	12 41 52.6	& +41 16 20.6	&	  8.20	& 3 	& SINGS/KINGFISH \\   
11	& NGC\,4647		     	&	12 43 32.6	& +11 34 53.9	& 	16.8~~~	& 3 	& HRS \\ 
	& NGC\,4649		     	&	12 43 40.0	& +11 33 09.8	&	17.3~~~   	& 3 	& HRS \\ 
12	& M51A			     	&	13 29 54.1	& +47 11 41.2	&	  7.69	& 3 	& SINGS/VNGS \\
	& M51B			     	&	13 29 59.7	& +47 15 58.5	&	  7.66	& 3 	& SINGS/VNGS \\
13	& NGC\,5394			&	13 58 33.7	& +37 27 14.4	& $56.4^{*}$~~ & 4 	& SHINING/GOALS \\
	& NGC\,5395			& 	13 58 37.6	& +37 25 41.2	& $56.4^{*}$~~ & 4 	& SHINING/GOALS \\
14	& M101			     	&	14 03 09.8	& +54 20 37.3 	&	  6.70	& 3 	& KINGFISH \\
	& NGC\,5474$^{+}$	     	&	14 05 01.2	& +53 39 11.6	&	  5.94	& 3 	& SINGS/KINGFISH 	
\enddata	
\tablecomments{Distance moduli were obtained from \citet{tully2008}, \citet{tully1988}, and the Extra-galactic Distance 
Database. Galaxies marked with $^{+}$ are dwarf galaxies with stellar mass of less than $1\times10^{9}$\,M$_{\odot}$. 
NGC\,2976/3077 and NGC\,4618\,4625 are dwarf pairs. The distances in Column 5
marked with $^{*}$ did not have distance moduli and were calculated based on heliocentric velocities, corrected per \citet{mould2000} 
and assuming \hubble. The determination of interaction stage is described in Section 2.2. In Column 6 we give the median of the Dopita system classifications. The surveys given in Column 7 include the Spitzer Infrared Nearby Galaxies Survey (SINGS), the Key Insights on Nearby Galaxies: a Far-Infrared Survey with Herschel (KINGFISH), the Herschel Reference Survey (HRS), the Very Nearby Galaxy Survey (VNGS), the Survey with Herschel of the ISM in Nearby INfrared Galaxies (SHINING), and the Great Observatories All-sky LIRG Survey (GOALS). }
\end{deluxetable*}

\subsection{Sample Selection}
Our galaxies are part of the Spitzer Interacting Galaxy Survey (SIGS)  (N. Brassington et al. 2014, in preparation).  The 103 galaxies of SIGS were selected strictly on the basis of interaction probability and hence cover a broad range of interaction stages. It is a sample of local galaxies because its selection criteria include a requirement that $cz~<4000~{\rm km~s^{-1}}$.

Paper I examines the fourteen systems with the most extensive wavelength coverage available at the time. This sample spans a range of interaction stages, having galaxies likely to be in their initial approach (e.g., NGC\,3424/3430) as well as galaxies in coalescence (e.g., NGC\,3690/IC\,694). It also covers a wide range of stellar masses (1$\times10^{8}-2\times 10^{11}~M_{\odot}$), stellar mass ratios (1:1 $-$ 1:40) and IR luminosities (3$\times10^{8}-8\times 10^{11}~L_{\odot}$). Figure \ref{par_range} shows the distribution of our sample's stellar mass and IR luminosity relative to the larger SIGS sample. Our sample covers most of the parameter space of the parent sample. Although consisting primarily of spiral-spiral interactions, our sample also contains two spiral-elliptical interactions. In Table \ref{sample}, we list our interacting galaxies along with distance and interaction stage estimates. In the systems with three or four galaxies, we will compare each of the three or six possible pairs with the simulations. While these more complex systems should ideally be compared to simulations of interacting groups, a pair-wise comparison provides a first step. These particular systems are sparse groups, which are not engaged in strong, multiple interactions, which would likely show stronger deviations from pair interactions than these poor groups.

\subsection{Photometry}
We assembled SEDs for each galaxy in our sample using publicly available photometry from the UV to FIR wavelengths. 
We measured global photometry in the larger of the two elliptical apertures necessary to contain the \emph{GALEX} near-UV (NUV) and \emph{Spitzer Space Telescope}'s \citep{wer04} Infrared Array Camera (IRAC) 3.6\um~emission. Here we summarize the available photometry in order of increasing wavelength.  

\begin{figure*}
\centerline{\includegraphics[width=0.85\linewidth]{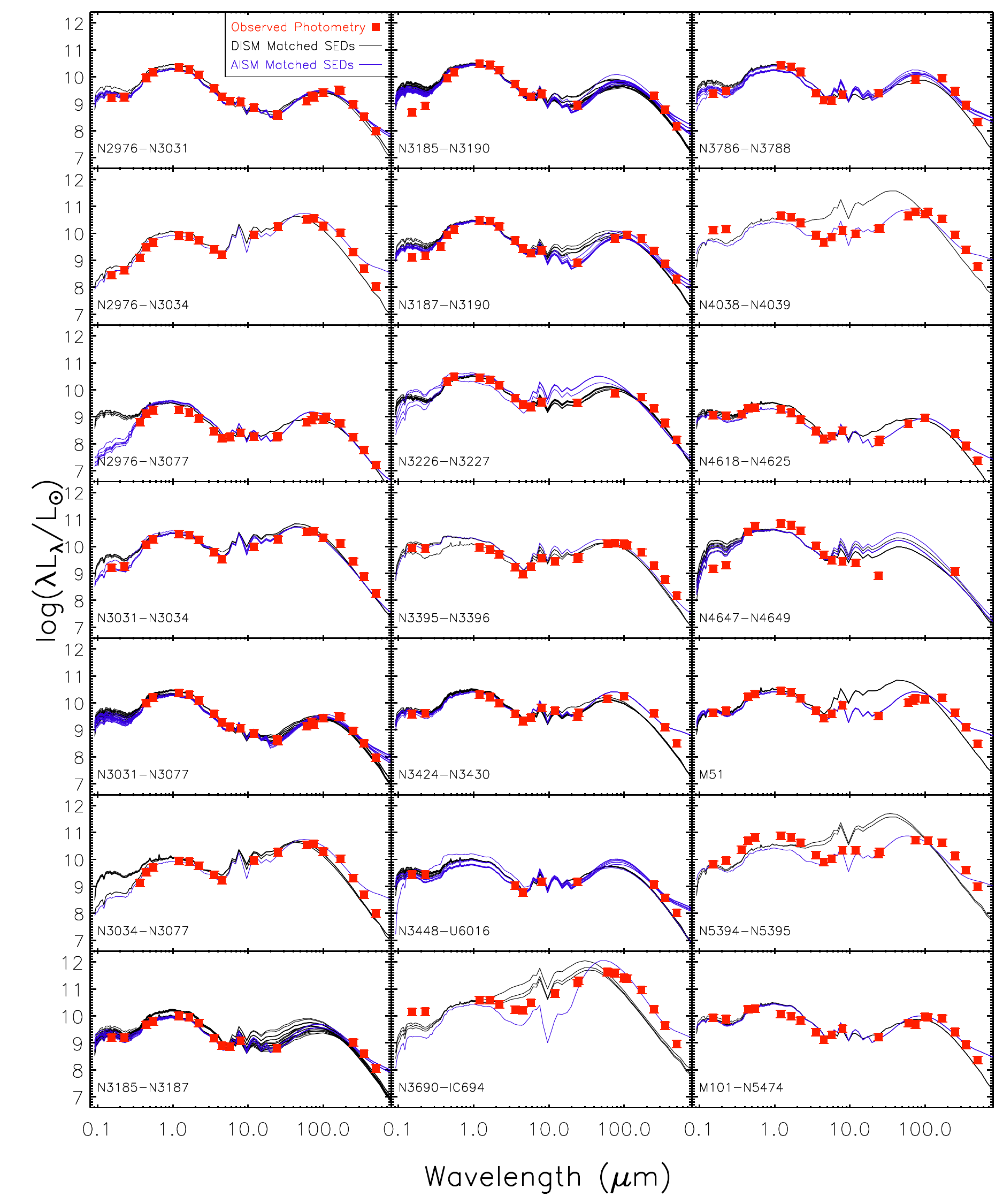}}
\caption{Best-matched simulated SEDs for the two treatments of the sub-resolution ISM structure, DISM (black lines) and AISM (blue lines), compared to the observed photometry (red) for the 21 pairs of interacting galaxies in our 14 systems. For most observed systems, at least one mock SED from the simulations provides a reasonably good (often statistically acceptable) fit to the observed SED, although there are a few cases for which the best matches are clearly unsatisfactory. The AISM SEDs often reproduce the SPIRE emission better. }
\label{int_fits}
\end{figure*}

\emph{GALEX} photometry was available for all but three of our galaxies (NGC\,3226, NGC\,3227, and NGC\,3077), which could not be observed  due to the presence of nearby foreground bright stars. Optical photometry were retrieved from the Third Reference Catalog \citep[RC3;][]{dev91}, which had \emph{UBV} for 50\% of the sample and \emph{BV} for an additional 25\%. The Two Micron All Sky Survey \citep[2MASS;][]{skru06} yielded near-IR (NIR) photometry for the whole sample. \emph{Spitzer}'s IRAC and Multiband Imaging Photometer (MIPS) instruments provided mid-IR (MIR) photometry from 3.6\um~to 24\um~for the whole sample. Measured photometry in the MIR was supplemented by ancillary photometry from \emph{Infrared Astronomical Satellite (IRAS)}  \citep{sur04, san03, soi89, mosh90}, and MIPS 70\um~and 160\um~data from SINGS \citep{dale05, dale07}. Lastly, FIR photometry was measured by PACS for twelve of fourteen systems and by SPIRE for all fourteen systems. Details of the photometry and their reduction can be found in Paper I. The photometry for each observed pair is shown as the red squares in Figure \ref{int_fits}. 

\subsection{Interaction Stage Classification}

Understanding galaxy interactions requires examination of systems at different interaction stages  as interactions proceed on timescales much too long for significant evolution to be observed in a single system. However, determining unambiguously where individual systems fall on the interaction sequence is not a straight forward process. For example, a pair of galaxies making their first close passage can appear very similar to a pair that has already passed near to each other and separated once more. Additionally, projection effects complicate the determination of the sequence of observed systems.

Here and in Paper I and Brassington et al. (in preparation),  we use the five-stage scheme devised by \citet{dopita02}. Stage 1 galaxies are non-interacting. Stage 2 galaxies have little or no morphological distortion. These systems are typically expected to be before or after the first passage. Stage 3 galaxies show a moderate degree of distortion, including tidal tails. Stage 4 galaxies show strong signs of disturbance and are expected to be in the more evolved interactions stages. Finally, the Stage 5 galaxies are post-merger systems. The systems considered in this work cover Stages 2 to 4. The simulations include both isolated and merging systems, and so span all five stages.

\subsection{Deriving Global Properties of the Observed Interacting Systems}

We use the SED fitting code MAGPHYS \citep{dac08} to estimate the SFR, sSFR, and stellar and dust masses. MAGPHYS fits SEDs with a stellar spectra library derived from the \citet{bru03} stellar population synthesis code and a thermal infrared dust spectrum. The ISM is modeled as a diffuse medium interspersed with denser stellar birth clouds. The dust emission is treated as the sum of four components:  two modified blackbodies of 30-60 K ($\beta=1.5$) dust and 15-25 K ($\beta=2$) dust, a MIR continuum consisting of the average of two $\beta=1$ modified blackbodies at 130 K and 250 K, and a polycyclic aromatic hydrocarbon (PAH) template \citep{madden06} with an 850 K ($\beta=1$) modified blackbody underlying continuum. MAGPHYS estimates galaxy  SFRs, stellar masses, dust masses, and dust temperatures. We provide MAGPHYS with the photometry in our set of 25 filters. We use a slightly modified version that provides SFR and sSFR estimates averaged over 1 Myr and 10 Myr, as well as the 100 Myr average that is output by the code by default.

\section{SIMULATIONS}

We based our analysis on a suite of hydrodynamic simulations of isolated and interacting galaxies for which we calculate the synthetic SEDs from UV to millimeter (mm) wavelengths using dust radiative transfer calculations\footnote{The interested reader can find animations of the evolution of the SEDs at \url{http://www.cfa.harvard.edu/sigs} and the full library of mock SEDs and auxiliary data about the simulations at \url{http://thedata.harvard.edu/dvn/dv/SIGS}.}. The methods we employed are described in detail elsewhere \citep[e.g.,][]{jon06, jon06b, jon10, hay11, hay12, hay13a, nar10b, nar10a}, so we only briefly summarize them here. The simulation suite includes four progenitor spiral galaxies that have properties similar to those of typical SDSS galaxies and span a stellar mass range from $6\times10^8$\,M$_{\odot}$ to $4\times10^{10}$\, M$_{\odot}$. These objects are referred to as M0, M1, M2, and M3 in Table  \ref{sims_desc}.  We simulated each progenitor in isolation (four simulations) and also performed binary galaxy merger simulations of each possible progenitor combination (ten simulations). 

At numerous times during each simulation, and from seven different viewing angles isotropically distributed in solid angle, we computed the emergent SEDs of the interacting and isolated systems. We compare the SEDs of our sample galaxies with the mock SEDs for all simulations, snapshots, and viewing angles. This comparison is the basis on which we assessed the simulation's ability to model the SEDs of realistic systems. 

\begin{deluxetable}{lrrrr} [bhp]
\tabletypesize{\scriptsize}
\tablecaption{Galaxy Models for the Simulations\label{sims_desc}}
\tablewidth{0 pt}
\tablecolumns{5}
\tablehead{
\colhead{} & \colhead{M3} & \colhead{M2} & \colhead{M1} & \colhead{M0} 
}
\startdata
M$_{*}~(10^{10}~$M$_{\odot}$)		&	4.22		&	1.18		&	0.38		& 	0.061 \\
Total Mass (10$^{10}~$M$_{\odot}$)	&	116.0~\,	&	51.0~\,	&	20.0~\,	&	5.0~~~~\\
M$_{{\rm Gas}}~(10^{10}~$M$_{\odot}$)	&	0.80		&	0.33		&	0.14		&	0.035 \\
Number of particles					& 240,000		& 150,000		&	95,000	&	51,000 \\
N$_{{\rm Dark~Matter}}$				& 120,000		& ~80,000		&	50,000	&	30,000 \\
N$_{{\rm Gas}}$					& ~50,000		& ~30,000		&	20,000	&	10,000 
\enddata
\tablecomments{Simulation parameters with further details given in  Tables 1 of \citet{jon06b} and \citet{cox08}. }
\end{deluxetable}

\subsection{Hydrodynamical Simulations}

We performed our suite of simulations of both isolated and merging galaxies using the TreeSPH \citep{hern89} code \gadgetthree \citep{spr05}, which uses a hierarchical tree method to compute gravitational interactions. The gas dynamics are solved via  smoothed-particle hydrodynamics \citep[SPH;][]{lucy77, gin77, spr10}, a pseudo-Lagrangian method that naturally yields higher resolution in denser regions.\footnote{Recent work \citep[e.g.,][]{age07, spr10b} has highlighted inherent inaccuracies in the SPH technique. Consequently, simulations performed using SPH can differ significantly from those performed using a more accurate moving-mesh approach \citep{vog12, ker12, sij12, tor12, bau12, nel13}. Fortunately, SPH is reliable for the types of idealized isolated disk galaxies and galaxy merger simulations presented here \citep{hay13b}.} 

To account for the unresolved structure of the ISM, the sub-resolution model of \citet{spr03}, which includes the effects of star formation and supernova feedback, is used. In this model, gas particles with density greater than a threshold of $n \sim 0.1$\,cm$^{-1}$ are assumed to follow an effective equation of state (EOS) that is stiffer than that for an isothermal gas. Gas particles that lie on the EOS form stars according to the volume-density-dependent Schmidt-Kennicutt law \citep{sch59, ken98}, SFR $\propto \rho_{\rm gas}^N$, with an index $N = 1.5$. Because SPH particles in our simulation have masses $\geq10^5$\,M$_{\odot}$, individual stars are not created. Instead, gas particles stochastically produce equal-mass star particles such that the SFR averaged over the simulation agrees with the rate given by the Schmidt-Kennicutt law. Black hole accretion and AGN feedback is included using the sub-resolution model of \citet{spr05b}.

Each model galaxy is composed of an exponential, rotationally-supported gas and stellar disk, a stellar bulge, and a dark matter halo; the latter two components are described using \citet{hern90} profiles. The progenitor disks are similar to the G0, G1, G2, and G3 models of \citet{jon06b} and \citet{cox08} except that the masses differ slightly. The galaxies are modeled to have median properties of SDSS galaxies and increase in mass from M0 ($6\times10^8$\,M$_{\odot}$ of stars) to M3 ($4\times10^{10}$\, M$_{\odot}$ of stars). We summarize the properties of these simulated galaxies in Table \ref{sims_desc}; all other properties are as given in \citet{jon06b} and \citet{cox08}. Figure \ref{par_range} shows how the ranges of the simulations' stellar mass and IR luminosity compare to those of the observed sample. Note that the range in stellar mass for a given simulation is rather small because the initial gas fractions are relatively modest and no additional gas is supplied to the galaxies during the course of the simulations. Although the simulations span the parameter space of the observed galaxies reasonably well, the coverage is not complete, and the sampling may be too coarse in some regions; if more simulations were performed to fill the gaps, the matches would likely be even better. 

We performed fourteen \gadgetthree simulations: one isolated simulation for each of the four progenitor galaxies and one merger simulation for each of the ten possible pair of galaxies. For the isolated simulations, each galaxy was allowed evolve secularly for 6\,Gyr. Because gas is not accreted from the surrounding environment in these idealized simulations, the SFR decreases as the gas is consumed. For the interactions, each pair of galaxies (M0M0, M1M0, M1M1, M2M0, M2M1, M2M2, M3M0, M3M1, M3M2, or M3M3) was placed on parabolic orbits such that the disks were prograde with initial separations increasing with the mass of the larger galaxy: 50\,kpc for M0, 80\,kpc for M1, 100\,kpc for M2, and 250\,kpc for M3. For simplicity, we used only one orbit, the e orbit of \citet{cox06}. Because this orbit is not `special' (i.e., the galaxies are not coplanar), this choice should not bias our results; still, it would be worthwhile to explore the effects of using multiple orbits in future work. Each pair was simulated  as it evolved from first approach through multiple pericenter passage to the final coalescence and post-merger stage. The  interactions take between 2.5\,Gyr and 6\,Gyr to reach the passively evolving stage at which we end a simulation.

\subsection{Radiative Transfer}

We used the 3-D polychromatic Monte Carlo dust radiative transfer code \sunrise \citep{jon06, jon10} to calculate spatially resolved UV-mm SEDs for the simulated galaxies at various times during the simulations and from different viewing perspectives. \sunrise calculates the emission from stars and AGN in the \gadgetthree simulation and performs radiative transfer to calculate the attenuation and re-emission from dust. {\sc Starburst99} \citep{leith99} SEDs are assigned to star particles, and the AGN particles emit the luminosity-dependent templates of \citet{hop07}. The dust distribution within each simulated galaxy is specified by the distribution of the gas-phase metals in the hydrodynamic simulation; a dust-to-metal density ratio of 0.4 \citep{dwek98, jam02} is used. For the purpose of the radiative transfer calculations, the dust density is projected onto a 3-D adaptive mesh. We use the Milky Way (MW) $R=3.1$ dust model of \citet{wei01} as updated by \citet[hereafter DL07]{drain07}. Dust temperatures are calculated assuming thermal equilibrium and depend on both the grain size and local radiation field.  The effect of dust self-absorption is accounted for using an iterative process. \sunrise calculates an SED per pixel, thereby yielding results analogous to integral field unit spectrography; however for this work, we only utilized integrated photometry. Seven viewing angles distributed isotropically in solid angle were used. Whereas the conditions of the hydrodynamic simulations are saved at 10\,Myr, the SEDs are typically calculated with \sunrise at 100\,Myr intervals. However, during the most active periods of the strongest interactions (i.e., when the SEDs vary rapidly in time), SEDs were calculated at 10\,Myr or 20\,Myr intervals.

For the simulations, the sub-resolution structure of the dust is a significant -- perhaps the most significant -- uncertainty in the radiative transfer calculations \citep[e.g.,][]{hay11, sny11, sny13, wuy09}.  When performing the radiative transfer through the dust in a simulation's ISM, \sunrise has two options for treating the sub-resolution dust structure: either the dust associated with the cold clouds in the \citet{spr03} sub-resolution model is ignored ($\rho_{\rm dust} = 0.4 \rho_{\rm metals,~diffuse}$; we refer to this as the `default ISM' or `DISM') or the total dust mass is used ($\rho_{\rm dust} = 0.4 (\rho_{\rm metals,~diffuse} + \rho_{\rm metals,~cold~clumps})$; we refer to this as the `alternate ISM' or `AISM'). In either case, to calculate the optical depth across a grid cell, \sunrise assumes that the dust mass ($\rho_{\rm dust}$) is distributed uniformly throughout the grid cell. Thus, the difference between the two ISM treatments is simply that in the DISM case, photons are propagated through less dust in each grid cell than in the AISM case. The effect of the alternative treatment varies between grid cells because the fraction of ISM contained in cold clouds depends on local ISM conditions. For each \gadgetthree simulation, we performed two \sunrise runs, one with each ISM model. Comparisons between the simulated and observed SEDs were done separately for each set, and we examined the effect of the ISM treatment on the selection of the best matches. 

The resulting suite of simulated SEDs of the fourteen simulations has 848 snapshots, each observed from seven viewing angles distributed isotropically in solid angle\footnote{The polar angle is sampled uniformly at cos($\theta$)=[$-1/3, 1/3, 1$] starting at the north pole and excluding the south pole. For each of these angles, the azimuthal angle is also uniformly sampled, except at the north pole where all azimuthal viewing angles are equivalent.} and run with both assumptions regarding the sub-resolution dust structure. Thus, our SED library contains a total of almost 12,000\,SEDs.

\begin{figure*}
\centerline{\includegraphics[width=0.65\linewidth]{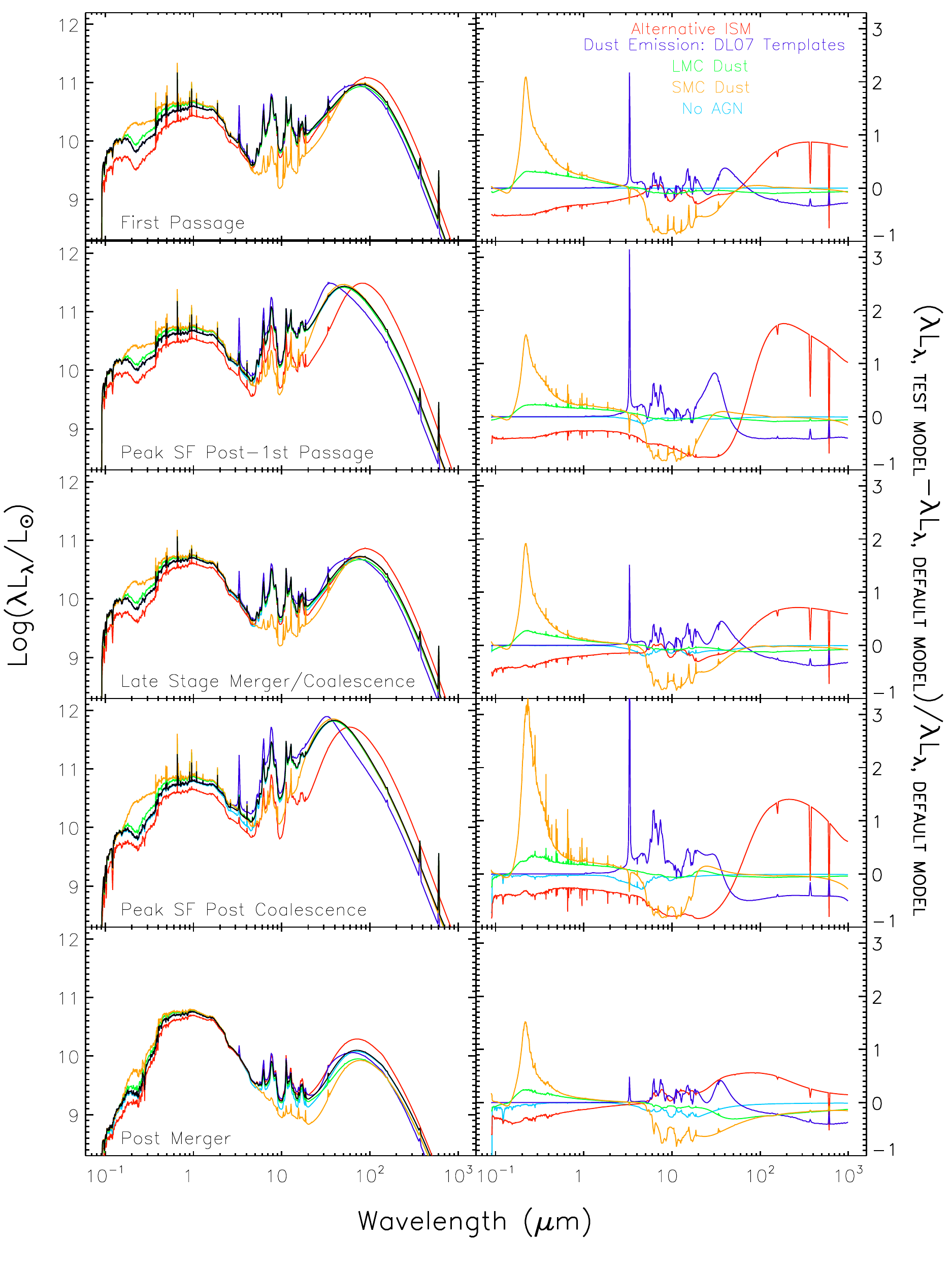}}
\caption{SEDs (left) for six different \sunrise radiative transfer runs at five times of interest (each row) for an equal-mass interaction similar to our simulations. The black line shows the SED of the default model, which assumes Milky Way-type dust and uses the DISM assumption (i.e., the dust in the cold phase of the sub-resolution ISM is ignored for the purposes of the radiative transfer calculation). In the right column, we show the fractional difference between each other model and this fiducial model. The red, alternative ISM SED shows the results when the radiative transfer is calculated using the AISM assumption (i.e., the total dust mass in a grid cell, rather than just the diffuse-phase dust mass, is used). The dark blue SED is the result of treating stochastically heated very small grain emission through the use of the Draine \& Li (2007) SED templates. The green and yellow SEDs are the result of assuming LMC- and SMC-type dust, respectively, rather than Milky Way-type dust. The cyan SED demonstrates the effect of removing the AGN contribution. The dust properties are the most significant uncertainty for the UV-optical region of the SEDs, whereas the ISM structure is most significant for the FIR because the AISM assumption yields more significant dust self-absorption and thus colder dust. In the MIR, multiple different model uncertainties are comparably important.}
\label{sunrise}
\end{figure*}

\begin{figure*}
\centerline{\includegraphics[width=\linewidth]{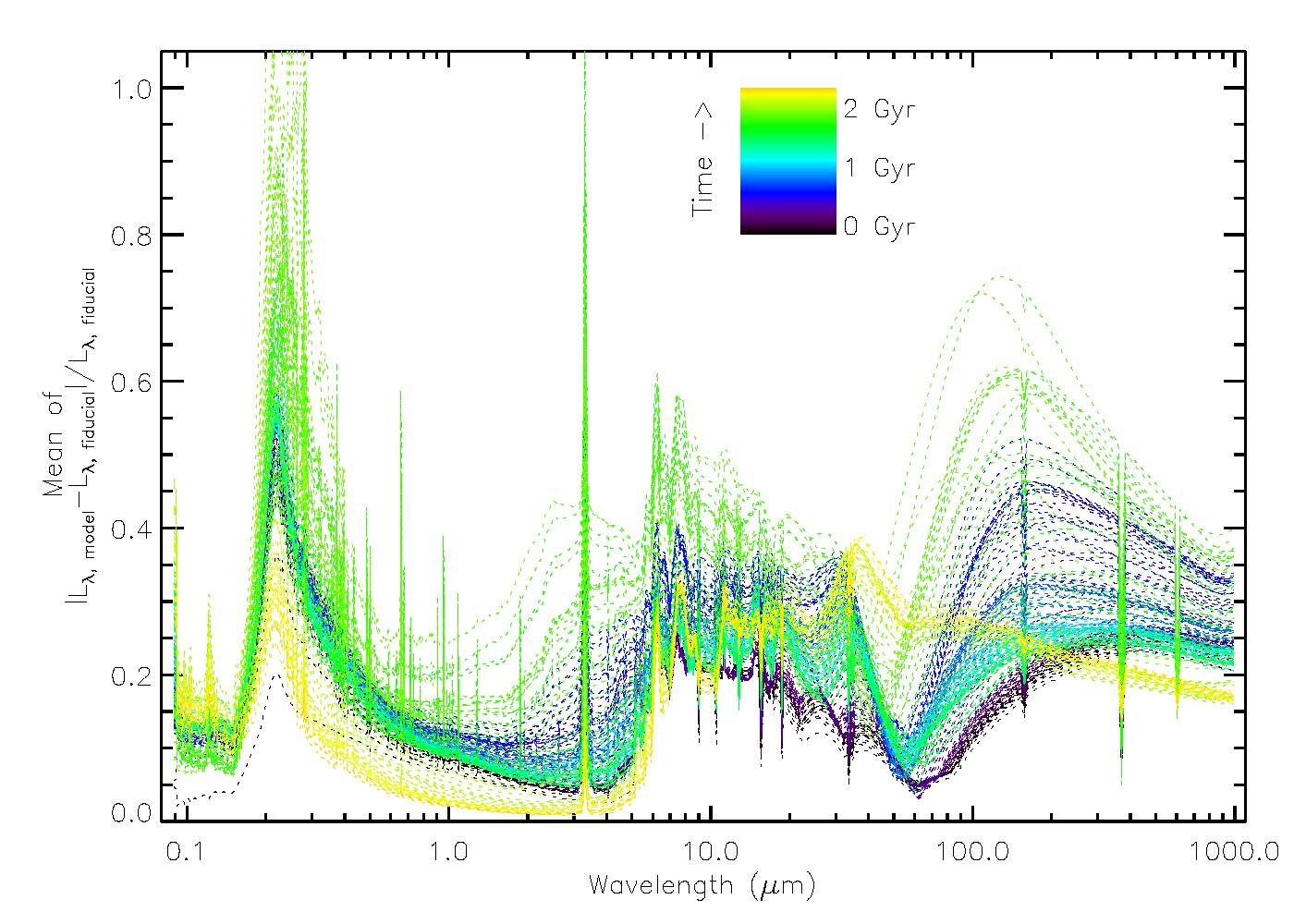}}
\caption{Each line shows the typical fractional difference as a function of wavelength due to assumptions made in the radiative transfer calculations for a single snapshot averaged over the five test models shown in Figure \ref{sunrise} and the seven viewing angles. Changing colors (from blue to green to yellow) show the evolution of time. For example, the variation in the FIR emission is typically about 30\%, but increases to 40$-$60\%  during the times that correspond to the second and fourth rows of Figure \ref{sunrise}, which corresponds to the starbursts induced near the first pericenter passage and final coalescence, respectively. This figure motivates the adoption of a uniform uncertainty of 30\% for the simulated photometry. }
\label{stdev_sunrise}
\end{figure*}

\subsubsection{Estimating the Uncertainty in the Simulated SEDs}

Radiative transfer codes inherently make assumptions about the material through which photons are propagated and the source of those photons. For example, in our \sunrise runs, we assumed MW dust composition rather than Large Magellanic Cloud (LMC) and Small Magellanic Cloud (SMC) dust compositions. In order to assess the uncertainty in the simulated SED due to the dust treatment, we examine six \sunrise runs calculated for an equal-mass spiral-spiral merger similar to our simulated interactions (the `weakly obscured' simulation of \citealt{sny13}), for which the hydrodynamic inputs remain constant but the assumptions used for the radiative transfer calculations were varied. Specifically, we varied the treatment of the sub-resolution dust structure, used two alternate dust models (LMC and SMC dust rather than the default MW model), used the DL07 dust emission templates which include the effects of stochastically heated very small grains instead of performing radiative transfer, and disabled the AGN emission.

 Figure \ref{sunrise} (left) shows SEDs at five different times during the interaction for the six different models. The black lne is the fiducial model. Figure \ref{sunrise} (right) shows the fractional difference between the fiducial model and each test model. As we noted, \sunrise has two possibilities for the treatment of the multiple ISM phases. The black fiducial model uses the default ISM treatment in which dense clumps are ignored. The red line in Figure \ref{sunrise} shows the SED derived when the alternative ISM methodology is employed and radiative transfer is calculated using the total dust content of a grid cell. The alternate ISM model has two main effects on the SED: colder dust temperatures (and hence enhanced emission in the SPIRE bands) and increased absorption in the optical and  UV. 

The green and yellow lines in Figure \ref{sunrise} show the SEDs that result with the assumption of LMC and SMC dust. Use of the SMC dust model results in significantly reduced NUV absorption (because of the lack of the 2175 \AA~feature in the SMC extinction curve) and weaker PAH features in the MIR (because of the reduced abundance of carbonaceous grains in the SMC dust model compared with the MW and LMC models). The results when the LMC dust model is used differ significantly less from the results for the MW model, but the attenuation in the UV-optical -- and thus the re-emission in the IR -- is somewhat less than in the default case. 

The results when the AGN emission is not included are denoted with cyan lines. The strongest effect is that compared with the default case the emission in the MIR at $\sim$10\,$\mu$m is reduced. This effect only becomes apparent in the later interaction stages when the AGN contribution to the SED is non-negligible.

Figure \ref{stdev_sunrise} quantifies how sensitive each wavelength band (from the UV to the FIR) is to the variations in the radiative transfer modeling. It elaborates on Figure \ref{sunrise} (right panels) which depicts that effect for a single snapshot and viewing angle for each of the five alternative models. Each line in Figure \ref{stdev_sunrise} shows the typical variation at a given snapshot, given by the median of $\frac{|(\lambda L_{\lambda})_{m, k} - (\lambda L_{\lambda})_{fiducial, k}|}{(\lambda L_{\lambda})_{fiducial, k}}$ over all test models $m$ and viewing angles $k$. To show the evolution of this uncertainty over the course the simulation, the color of the line varies with snapshot from purple to blue to green to yellow. We find that the uncertainty in the NIR and MIR exhibits little evolution with time. In these wavelength regimes, the variation with respect to the fiducial model is typically $\sim$20\% and 30$-$40\%, respectively. The MIR is  dominated by the reduced PAH emission for the SMC dust model due to the decreased abundance of carbonaceous grains and the increased emission from stochastically heated grains when the DL07 templates are used. The significantly lower absorption in the NUV for the SMC model results in the high standard deviation around 0.2-0.3\,$\mu$m. The standard deviation in the FIR is dominated by the assumption regarding the sub-resolution dust clumpiness and is typically at least 40\%.The FIR variation also exhibits the most evolution with time. Its standard deviation rises from 40\% to $\sim$80\% during first passage and to 100\% during coalescence. The median uncertainty taken over all the snapshots and over the whole SED is 30\% for a single viewing angle and 35\% overall.

\begin{figure*}
\centerline{\includegraphics[width=0.82\linewidth]{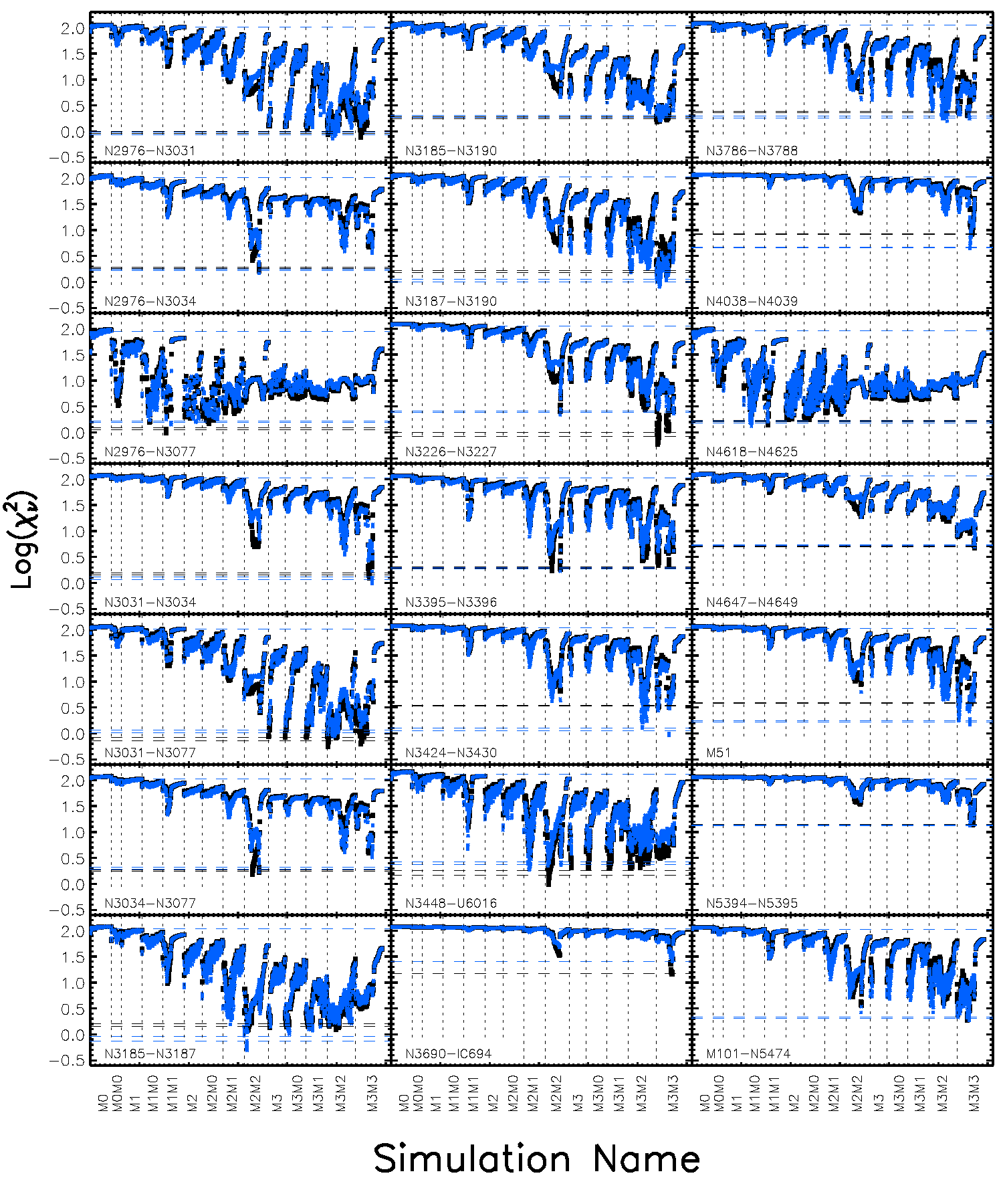}}
\caption{Reduced $\chi^2$ in the comparison of each simulated and observed SED. The default and alternative ISM SEDs are indicated with black and blue points, respectively. The 14 simulations are labeled on the horizontal axis and are separated by vertical dashed lines in each panel. The horizontal lines in each plot show the selection limits for the two sets of best matches (lower lines) and the set of worst matches (upper lines) in the same color schemes as the points. These lines show that some galaxies (e.g., NGC\,3424/3430 and M51) are better matched with the AISM SEDs, others are better matched with the DISM SEDs (e.g., NGC\,3226/3227 and NGC\,3031/3077), and others are similarly well matched (e.g., NGC\,3395/3396 and M101/NGC\,5474). }
\label{chi_by_model}
\end{figure*}

\section{METHODOLOGY}

\subsection{Matching Criterion}

We seek to determine whether the SED of an observed system is well-matched by one or more simulated SEDS, and, if there are any good matches, whether they come from a small region of the simulation parameter space (and thus the matching procedure yields a non-degenerate `fit'). We therefore chose to do a brute-force evaluation of all simulated SEDs with the SED of each interacting system by means of the  $\chi^2$ statistic between each pair of simulated and observed SEDs:

\begin{eqnarray}
\chi^{2} =   \sum_{{\rm SED}}  \frac{{\rm (L_{\nu, Data}-L_{\nu, Model})^2}}{{\rm \sigma_{Data}^{2}+\sigma_{Model}^{2}}}  
\end{eqnarray}

As discussed in Section 3.2, there is significant uncertainty in the models, which is primarily due to uncertainties regarding the dust properties and the need to treat sub-resolution dust structure in the simulations in a simplified manner. Based on the analysis described in Section 3.2.1,  we determined that a uncertainty of 30\% realistically represented our confidence in the simulated SEDs. The uncertainty in the observed photometry is primarily driven by the calibration uncertainty of the instruments, which is typically around 10\%. Therefore, the statistic we use to compare the observed and the model SEDs is:

\begin{eqnarray}
\chi^{2} =   \sum_{{\rm SED}}  \frac{{\rm (L_{\nu, Data}-L_{\nu, Model})^2}}{{\rm (0.10\times L_{\nu, Data})^{2}+(0.30\times L_{\nu, Model})^{2}}}  
\end{eqnarray}

\subsection{Selection of the Best and Worst Matches}

In Figure \ref{chi_by_model}, we show the reduced\footnote{Because we examine the trends as a function of simulation, snapshot (or time), and viewing angle, we effectively have three free parameters.} $\chi^2$ value for each pairing of an observed and simulated SED ordered by simulation. The horizontal lines in this figure identify the sets of best matching simulated SEDs that we will examine for trends: those within $\Delta \chi^2 \leq 3$ and within $\Delta \chi^2 \leq 5$ of the minimum $\chi^2$ for each observed system. This selection was done separately for the DISM and AISM SEDs. The smaller set of matches on average has 9 matches per observed pair for the DISM SEDs and 7 matches for the AISM SEDs.  The larger set of matches provides a sense of the stability of the trends if we relax our definition of the best matches, and on average has 19 and 13 matches per observed pair for the DISM and AISM SEDs respectively. We also use the match criterion to select the worst matches. We select the set for each interaction that cover 10\% of the $\chi^2$ range and have the largest  $\chi^2$ values, typically $\sim$1000 simulated SEDs.

For each observed system, we also determine the mean and median $\chi^2_{\nu}$ as a function of simulation, snapshot, and viewing angle, to determine whether broad areas of parameter space can be deemed unlikely to reproduce the observed SED for each set of \sunrise runs. We describe the trends in the matches and in these functions in Section 5.1.

\subsubsection{Best-Matched SEDs}

In Figure \ref{int_fits}, we plot the best-matched simulated SEDs for each interaction overlaid with the observed photometry. Typically, the observed systems can be matched reasonably well by one or more of the simulated SEDs, although the `fits' would not always be considered acceptable in a statistical sense. The success of the simulated SEDs at reproducing those of the observed systems is encouraging because we are self-consistently `forward-modeling' the SEDs using dust radiative transfer performed on hydrodynamical simulations and not tuning any parameters. We stress that because we do not allow the normalization of the simulated SEDs to be free (i.e., we are not using them as templates that can be arbitrarily rescaled), the SEDs are intimately tied to the physical parameters of the simulations. The only manner in which we can modify the outcome of the fitting procedure is by performing additional simulations or by varying the assumptions in the hydrodynamical simulations and radiative transfer. Here, we have done the latter for the most significant uncertainty in the radiative transfer calculations, the sub-resolution dust structure, and we will discuss the effects of varying this assumption below. 

Several systems show interesting behavior. Some of our most evolved systems (e.g. NGC\,4038/4039, and NGC\,5394/5395) have better overall matched SEDs from the AISM set than from the DISM set, although their photometry hints at the presence of typically cooler dust than found in the simulations as their FIR emission peaks at longer wavelengths. These systems also typically have excess absorption in the UV relative to the observed photometry, and have very few matches. In systems where the DISM set yields better overall matches (e.g. NGC\,3226/3227, NGC\,3690/IC\,694), the AISM model reproduces the observed FIR emission as well or better than the DISM model, but the MIR emission differs significantly.

The UV emission shows the greatest degree of variation between best matches, particularly when unconstrained by observations (e.g. NGC\,2976/3077). In more distorted systems (e.g. NGC\,3690/IC\,694), its absorption is overestimated. The two pairs that include the large edge-on heavily obscured spiral NGC\,3190, in contrast, did not find matches with sufficient absorption, and the UV emission of the pair containing the large elliptical NGC\,4649 is likewise overestimated.

\subsection{Determination of Simulation Parameters for Comparison}

Having established which simulated SEDs were best matched to each observed system, we compare the associated physical parameters such as IR luminosity, stellar mass, dust mass, and SFR to estimate how accurately they are recovered. Many parameters are tracked by the hydrodynamic simulation (e.g. stellar mass and SFR). The dust mass is assumed to be comprised of 40\% of the metals within the gas (also tracked during the simulations). The 3-1000\,\um~luminosity is calculated as part of the \sunrise post-processing. We also compare the dust temperature estimates. To do so, we calculate a representative temperature by fitting the available FIR ($\lambda \geq 50\mu$m) photometric data points  of both the observed and simulated SEDs with a single $\beta$=1.5 optically-thin modified blackbody model. Since some of our observed systems only have SPIRE data in the FIR, we were restricted to two free parameters. We also fit a similar blackbody with $\beta$=2 to estimate the uncertainty due to fixing $\beta$.

\section{Discussion }

\subsection{Where Do the Best-Matched SEDs Come From?  }

In any comparison of a suite of models to observations, two of the most important questions asked are: which areas of parameter space can be ruled out and which areas of parameter space give us the best matches?  For the sets of best and worst matches for the comparisons with the DISM and AISM SEDs, we examine trends in the distributions of simulations, snapshots, and viewing angles from which the matches originate. For each parameter of interest, we first examine the behavior of the mean and median $\chi^2_{\nu}$ as a function of the parameter and then discuss the source distributions.

\subsubsection{Results with Model SEDs Using Default ISM}

We will first examine the trends in the origins of the best and worst matched DISM SEDs. The DISM SEDs provide good matches for some of the observed systems (e.g. NGC\,3226/3227), although, for others (e.g. NGC\,3424/3430), the best-matched DISM SEDs clearly underestimate the SPIRE emission as can be seen in Figure \ref{int_fits}. We find that the best matches typically come from the same simulation, often from periods near coalescence, but generally do not have a preferred viewing angle.

\begin{figure*}[thp]
\centerline{\includegraphics[width=\linewidth]{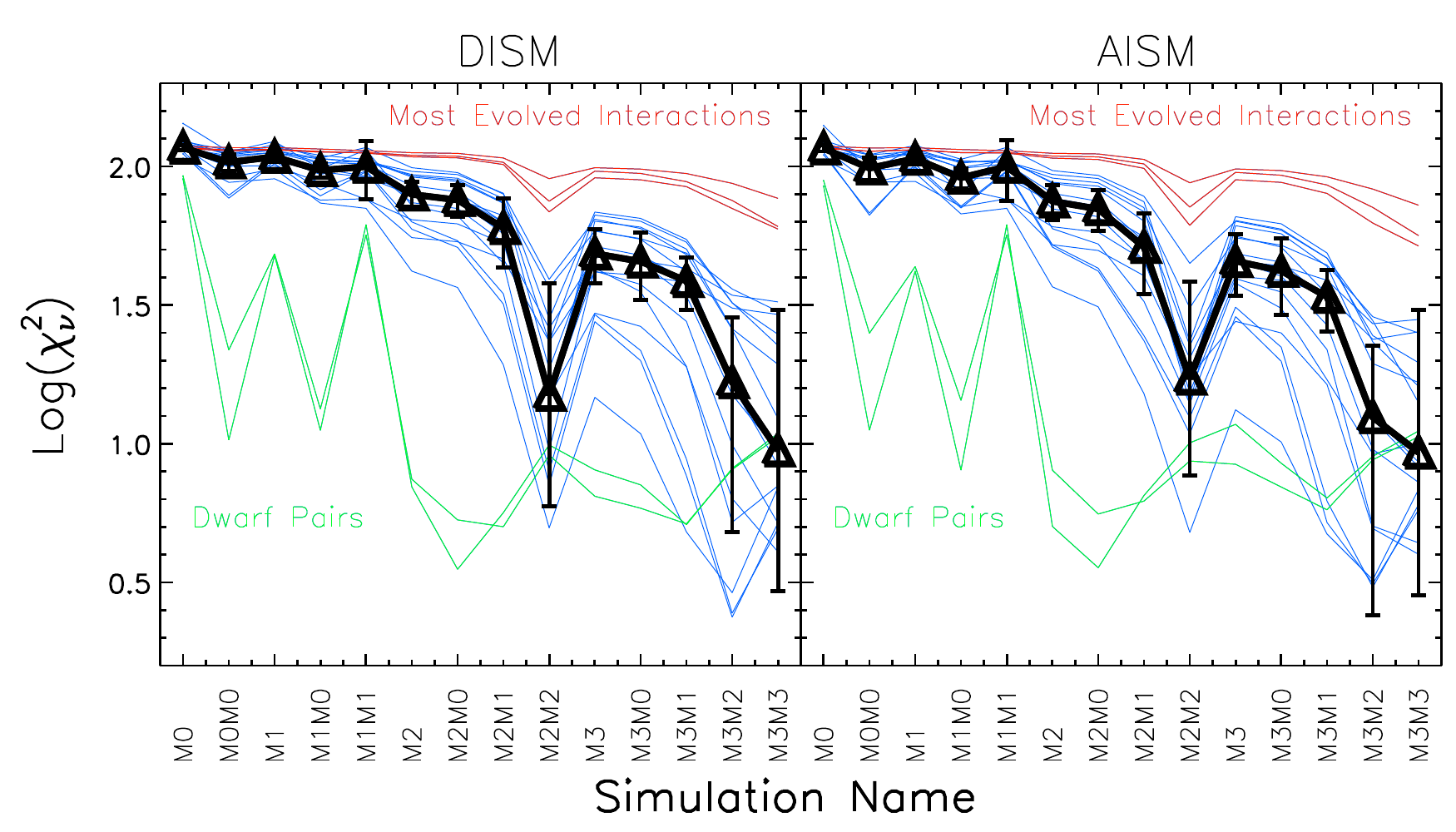}}
\caption{Median $\chi^2_{\nu}$ as a function of simulation for each interaction. The black triangles is the median of the blue lines, which all have similar tendencies, showing minima indicating increased likelihood at M2M2 and at the massive end of the simulations at M3M3 and M3M2. The three lines in red are NGC\,3690/IC\,694, NGC\,4038/4039, and NGC\,5394/5395, our most evolved interactions which are generally least well-matched to these simulations and hence have flatter distributions. The two green lines are the two dwarf pairs, NGC\,2976/3077 and NGC\,4618/4625, which have much more variable functions, whose minimum is around M2M0 and M2M1.}
\label{chis_by_sim}
\end{figure*}

\begin{figure*}
\centerline{\includegraphics[width=0.9\linewidth]{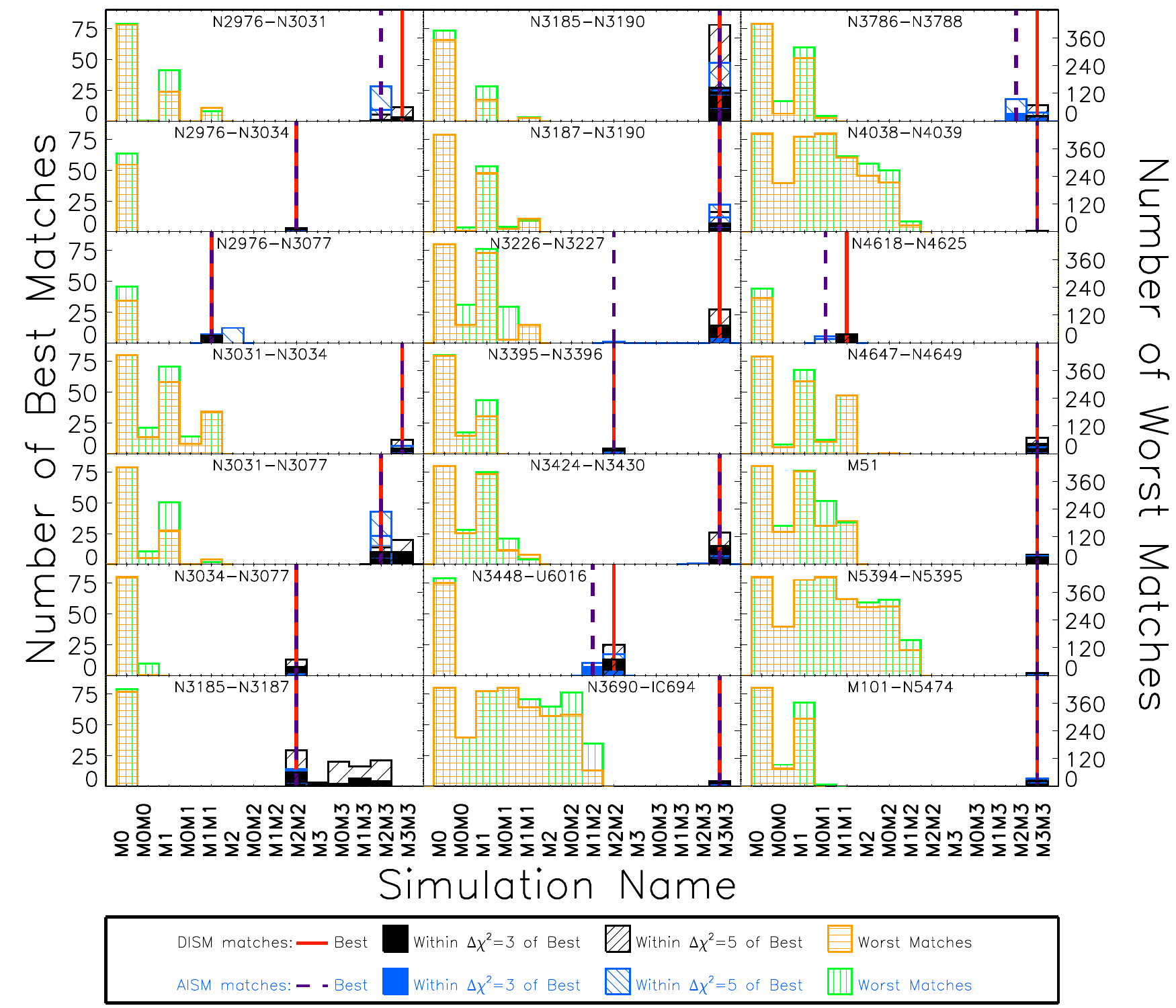}}
\caption{Distribution of the simulations from which the best matches to the DISM (black) and AISM (blue) mock SEDs compared to the origins of the worst matches (green and yellow respectively). The smaller set of best matches (within $\Delta \chi^2$=5 of the best match) is filled in over the hashed distribution of the larger set of best matches (within $\Delta \chi^2$=3 of the best match). The best single match from each set is shown with the red (DISM) and purple (AISM) lines; in many cases, the best-matching simulation is independent of the sub-resolution ISM model used. The best matches typically come from more massive interaction simulations and generally originate from 1-2 simulations, whereas the worst matches come from simulations of less massive galaxies, particularly the isolated simulations of M0 and M1.}
\label{int_sim}
\end{figure*}

\paragraph{Matches as a Function of Simulation}

Figure \ref{chis_by_sim} shows $\chi^2_\nu$ as a function of simulation ordered by increasing mass from M0 to M3M3. We determined the median $\chi^2_\nu$ over all of the viewing angles and time-steps  for each simulation as a means of identifying regions in the simulation parameter space that contain the best and worst matches. We find similar behavior for 16 of our 21 interactions, which show increased likelihood of matching the M2M2, M3M2, and M3M3 SEDs. The three most evolved systems have much flatter likelihood functions, but also hint at similar behavior. In contrast, the two dwarf pairs (NGC\,2976/3077 and NGC\,4618/4625) have much more varied distributions.

Although Figure \ref{chis_by_sim} shows the likelihood distribution as a function of simulation, it does not clearly show the variety in the number of matches or their distribution between simulations. For each observed interaction, we plot in Figure \ref{int_sim} the distribution of the best (black) and worst (green) matched DISM SEDs (determined as described in Section 4.2). The best matches typically come from the same simulation, which is very rarely an isolated galaxy simulation (the only one is NGC\,3185/3187).  Most of the best matches come from the M3M3, M2M2, and M3M2 simulations.  The worst matches always originate from less massive simulations and their distributions are often dominated by the isolated M0 and M1 simulations. 

Figure \ref{int_sim} also demonstrates another tendency. Systems with very few good matches (e.g. NGC\,3690/IC\,694, NGC\,4038/4039, NGC\,5394/5395) also have the largest number of worst matches, many of which originate from the interactions between the three less massive simulated galaxies. The $\chi^2$ distributions for these galaxies (see Figure \ref{chi_by_model}) is flatter than those of other systems, particularly at the low-mass (M0) end, indicating many similarly bad matches. Many of these models are not representative of these three systems in part due to a significant difference in the stellar mass, which has a broad normalizing effect on SEDs. Further, these three systems are among our most evolved and there are only a few snapshots in each interaction simulation that capture the coalescence in its most active phases during which the SED evolves rapidly and variation with viewing angles can become significant because the optical depth of the central starburst can be very high.

Figures \ref{chis_by_sim}$-$\ref{int_sim} together demonstrate the general trends of matches with simulations: (1) the best matches to any given system typically come from only one simulation; (2) the most massive major merger models generally yield the most best matches for our observational sample, while the simulations of less massive galaxies generally result in the worst matches; (3) despite the range in mass ratios in the observed systems, only the pairings of NGC\,3031/3077, NGC\,3185/3187, and NGC\,2976/3031 have some of their best matches to DISM originating from a non-equal-mass merger; and (4) our most evolved and massive interactions (e.g., NGC\,3690/IC\,694, NGC\,4038/4039, NGC\,5394/5395) have the fewest good matches.

\begin{figure*}
\centerline{\includegraphics[width=\linewidth]{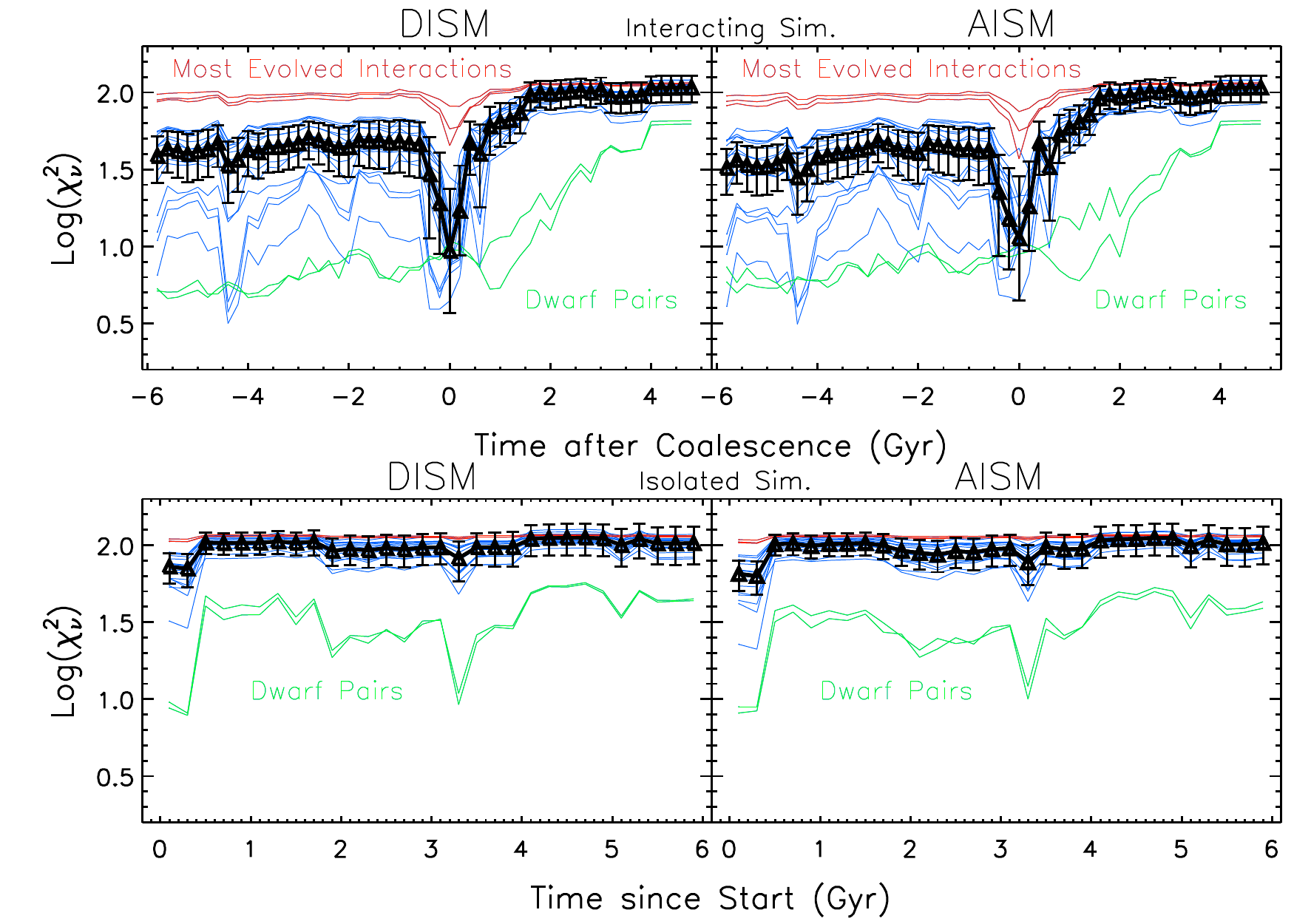}}
\caption{Median $\chi^2_{\nu}$ as a function of time relative to the time of coalescence (interactions, top panels) or the beginning of the simulation (isolated simulations, bottom panels). The black triangles is the median of the 16/21 systems (blue lines), which all have similar tendencies, showing approximately constant likelihood with lower $\chi^2_{\nu}$ in the $\sim$1\,Gyr before and after coalescence. Three systems (NGC\,2976/3031, NGC\,3031/3077, and NGC\,3185/3187) also show greater likelihood around 4\,Gyr before coalescence. As in Figure \ref{chis_by_sim}, the three evolved systems (red) have flatter distributions, but they also show increased likelihood around coalescence. In contrast, the the two dwarf pairs (green) show increasing $\chi^2_{\nu}$ with time in the interacting simulations and have more variable functions of $\chi^2_{\nu}$ with time in the isolated simulations.}
\label{chis_by_snp}
\end{figure*}

\begin{figure*}
\centerline{\includegraphics[width=0.9\linewidth]{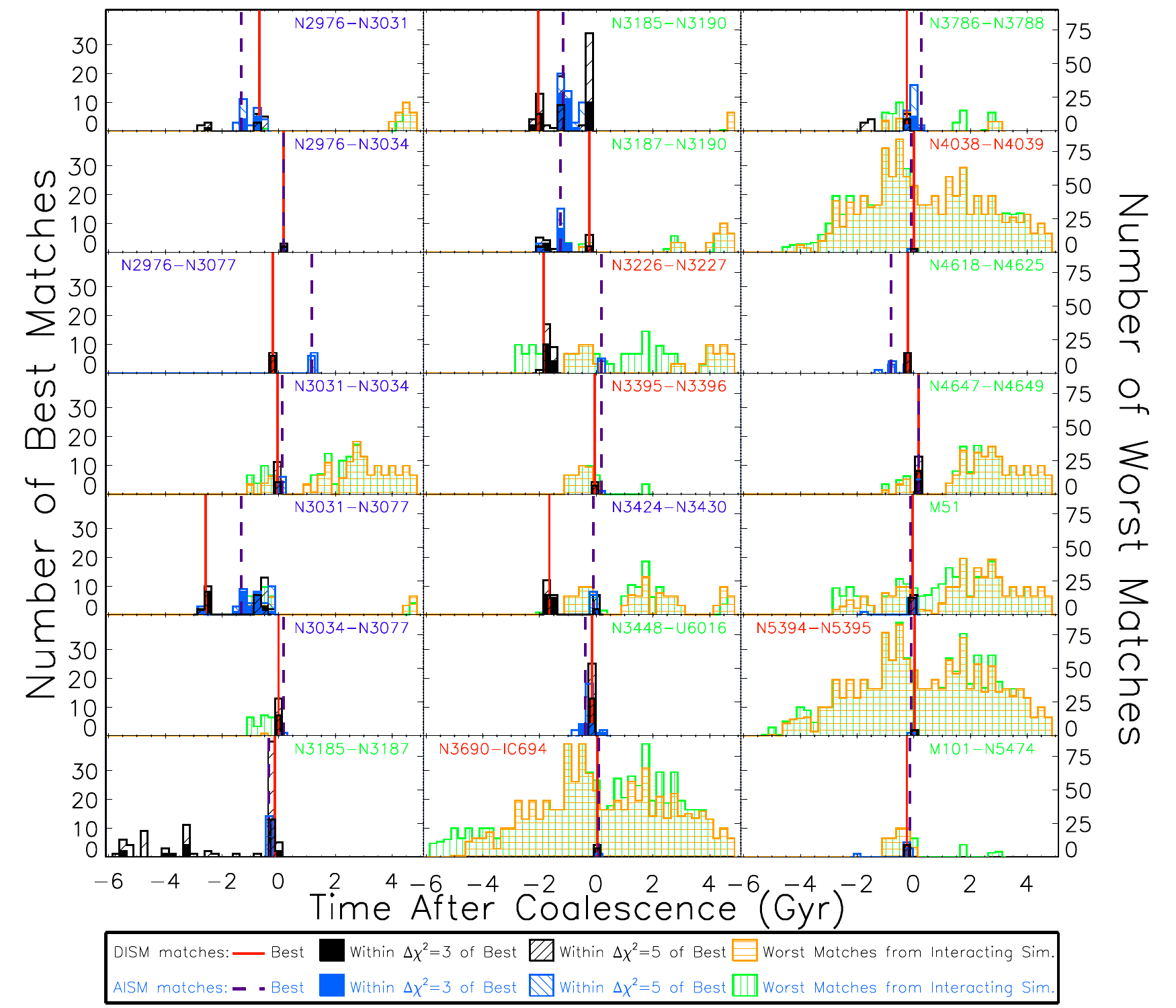}}
\caption{Distribution of the times to coalescence in 200\,Myr intervals of the best and worst matches using the same color scheme as Figure \ref{int_sim}. The color of the name indicates weakly (blue), moderately (green), and strongly (red) interacting systems based on the Dopita et al. (2002) classification system (\S2.3). We do not show the matches originating from isolated galaxy simulations, since the time to coalescence would not be definable; therefore some systems do not have any plotted worst matches. In the cases for which the distribution of best-matching times is narrow, as is the case for most of the strongly interacting systems and some of the others, the SED comparison is able to infer information about the physical state of the system; in the other cases, additional information is necessary. Note that in many cases, changing the sub-resolution ISM treatment does not significantly alter the best-matching times. Note that many of the worst matches come from significantly after coalescence, but the bulk of the best matches come from close to coalescence.  }
\label{int_snp}
\end{figure*}

\paragraph{Matches as a Function of Time}

In Figure \ref{chis_by_snp} we show the the median $\chi^2_\nu$ as a function of time relative to coalescence of the supermassive black holes (SMBH) for interactions and time since the simulation start for the isolated galaxies. As in Figure \ref{chis_by_sim}, we find a similar behavior in most of observed systems of increased likelihood of a match with an interaction in the one Gyr before and after coalescence. The most evolved systems likewise show only slightly increased likelihood near coalescence. The dwarf pairs generally show decreasing likelihood with time. Three other systems (NGC\,2976/3031, NGC\,3031/3077, and NGC\,3185/3187), which each have a low-mass component\footnote{NGC\,2976, NGC\,3077, and NGC\,3187 have stellar masses of $1\times10^{9}$\,M$_{\odot}$, $7\times10^{8}$\,M$_{\odot}$, and $3\times10^{9}$\,M$_{\odot}$, respectively (Paper I).}, also have increased likelihood $\sim$4\,Gyr prior to coalescence. In contrast, the isolated simulations remain approximately equally unlikely to match over their entire duration, except for the dwarf pairs which have a more variable function of $\chi^2_\nu$ with time.

Figure \ref{int_snp} shows the distribution of the times from which the best and worst matches from the interacting simulations originate. NGC\,3185/3187 is the only one to match with M3 and its three best matches occur 400 Myr after the simulation start. For most systems, the best matches occur in the Gyr before and after coalescence. However, a few systems have best matches from earlier in the simulation (e.g. NGC\,3031/3077, NGC\,3185/3190, NGC3226/3227). NGC\,3185/3187 in particular has a wide distribution of best-matched times. We note that we do not see a correlation between the \citet{dopita02} interaction stage and the times of the best matches (i.e. ``strongly interacting'' systems (red) do not necessarily have best matches from later in the simulations than ``weakly interacting'' systems (blue)).

In conclusion, we find that: (1) best matches to our observed systems often cluster around coalescence and primarily populate times before coalescence, while (2) the worst matches to our set of observations, in contrast, generally originate in the post-merger interaction stages (or are confined to simulations of isolated galaxies). 

\begin{figure*}
\centerline{\includegraphics[width=0.9\linewidth]{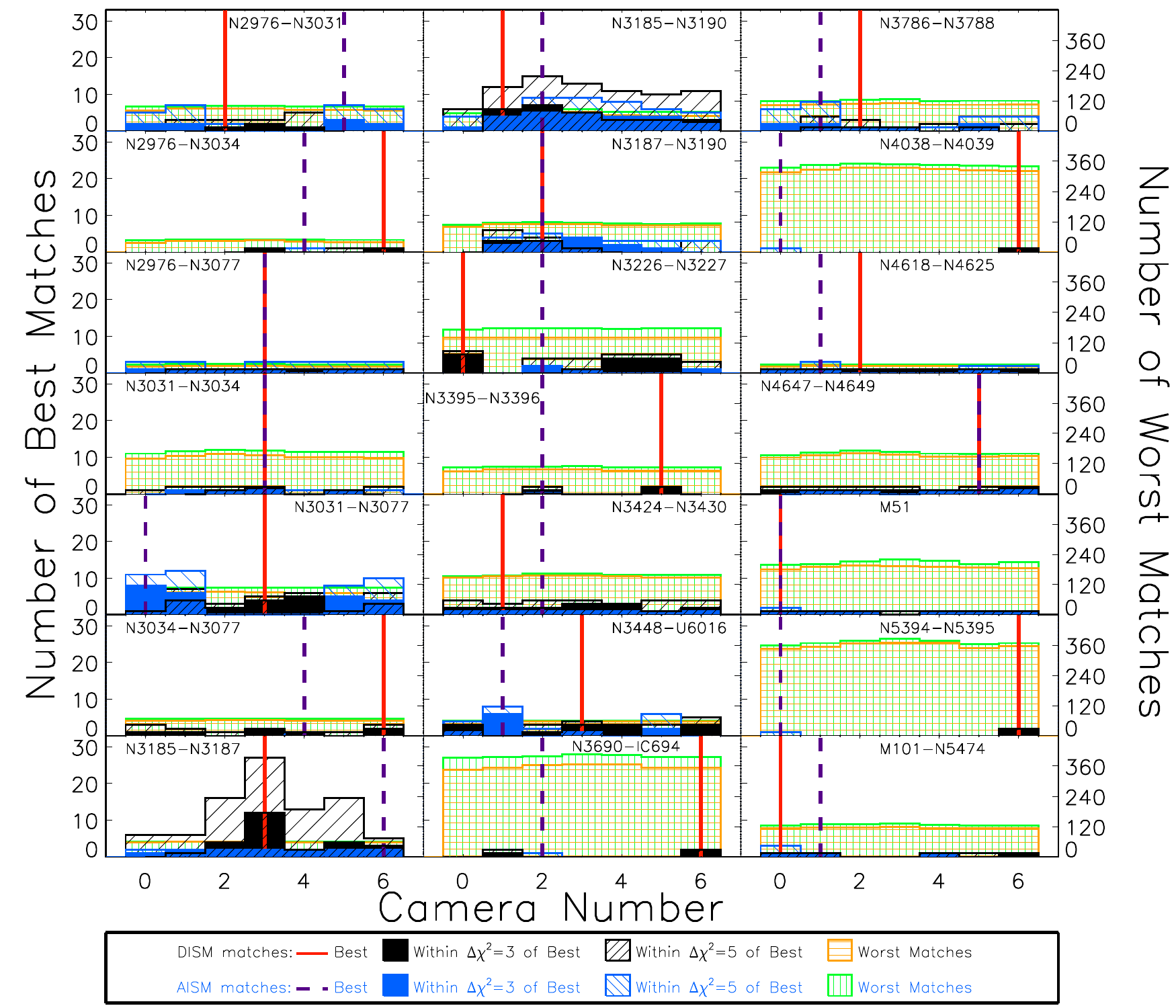}}
\caption{Distribution of the viewing angles of the best and worst matches using the same color scheme as Figure \ref{int_sim}. Typically, no viewing angle is preferred, which is expected because none of the viewing angles are `special' (e.g., none are along axes). The only exception is N3185-N3187, which exhibits  a preference for viewing angle 3 for its best matches; this preference may exist because the best matches come from significantly before coalescence (see Figure \ref{int_snp}), at which time the spirals' disks are largely intact and therefore inclination affects the derived SEDs significantly.}
\label{int_cam}
\end{figure*}

\paragraph{Matches as a Function of Viewing Angle}

The median $\chi^2_\nu$ as a function of viewing angle were uniformly flat, so we do not show a figure for the viewing angle analogous to Figures \ref{chis_by_sim} and \ref{chis_by_snp}. Figure \ref{int_cam} shows the distribution of the viewing angles of the best and worst matches, which are also fairly uniform. Since merging spirals are angled with respect to one another, there is no special viewing angle that yields both galaxies edge-on or face-on. Further, once an interaction has disrupted the disks, the SEDs from the different viewing angles become increasingly similar. The only exception is NGC\,3185/3187, whose DISM matches, show a distinct preference for Camera 3. This is likely due to matches from early in the simulated interactions (see Figure \ref{int_snp}) at which time the galactic disks are not yet fully disrupted.

\subsubsection{Results with Model SEDs Using Alternative ISM}

We performed a second set of \sunrise runs with an alternative treatment of the ISM, which better matches the FIR emission of some observed systems.  In this section, we discuss how the origins of the matches from the alternative ISM model SEDs differ from those of the DISM SEDs. We find that the DISM and AISM SEDs are similarly well-matched for many systems, although some systems (e.g. NGC\,3424/3430) are better matched to the AISM SEDs. The trends with simulation and viewing angle for the matches with the AISM SEDs are similar to those with the DISM SEDs, but we do see some differences in the time of the best matches. We see little difference in the distributions of viewing angle for the DISM and AISM matches.

\paragraph{Matches as a Function of Simulation}

The AISM panel of Figure \ref{chis_by_sim} is very similar to the DISM panel. Likewise, many of the systems have similar distributions of the simulations from which the best (blue) and worst (yellow) matches originate (Figure \ref{int_sim}). Seven systems show some differences in their distributions. NGC\,2976/3031,  NGC3448/UGC\,6016, NGC\,3786/3788, and NGC\,4618/4625 have best-matched AISM SEDs from a simulation of less massive galaxies than the DISM matches (e.g. M3M2 rather than M3M3). NGC\.2976/3077 has more than half of its AISM matches from the isolated M2 simulation. Figure \ref{int_fits} shows the drastic difference in the UV emission of the best-matched AISM and DISM SEDs. This wavelength range remains unconstrained because  \emph{GALEX} have observed NGC\,3077 due to a nearby bright star. NGC\,3226/3227, likewise without UV data, has a similar difference, although with an additional best match at M2M2 instead.  NGC\,3185/3187 has the opposite difference between its DISM and AISM matches, with a much smaller number of matches to the AISM SEDs.

\paragraph{Matches as a Function of Time}

As with Figure \ref{chis_by_sim}, there is little difference between the AISM and DISM panels of Figure \ref{chis_by_snp}.
Figure \ref{int_snp} shows more differences between the AISM and DISM matches, although many systems also have matches clustered near coalescence. Three systems (NGC\,2976/3031, NGC\,3187/3190, and NGC\,4618/4625) tend to match to earlier times in the AISM SED set than in the DISM SED set. Six systems, however, tend to match to later times for the AISM SEDs than the DISM SEDs. The other main difference between the AISM and the DISM SEDs is the distribution for NGC\,3185/3187, which is much narrower for the AISM matches. \\

In conclusion, we find that varying the ISM, while it makes distinct differences in some wavelength bands does not systematically improve the overall matches as measured by $\chi^2$.

\subsection{Recovery of Global Galaxy Properties }

Having determined which simulated SEDs were best matched to each observed system, we estimate how accurately these simulations recover physical parameters such as IR luminosity, stellar mass, dust mass, and SFR from the SEDs. To that end, we compare the best-matched simulations to the quantities derived using the MAGPHYS code; because MAGPHYS performs SED modeling rather than predicting SEDs from simulations, it is considerably more flexible and can yield good fits to the observed SEDs. Thus, MAGPHYS can estimate the physical properties of the observed systems for comparison with those of the simulations. In Figure \ref{par_comp1} and \ref{par_comp2} we plot the parameters associated with the best-matched simulated SEDs against the MAGPHYS-derived value for the observed systems. We also compare the typical temperatures of the best matches from the blackbody fits.

\subsubsection{IR luminosity}
We find that the IR luminosity (left panels of Figure \ref{par_comp1}) is well recovered, as expected because the simulation SEDs typically match the observed IR SEDs reasonably well. The AISM matches are a better match for some of our most evolved systems (i.e., NGC\,4038/4039 and NGC\,5394/5395). The most evolved systems are also those with the highest IR luminosity. The Stage 2 (blue diamonds) pairs with IR luminosities of $\sim10^{11}$\,L$_{\odot}$ are the three pairs with NGC\,3034.

\subsubsection{Star Formation Rate}

\begin{figure*}
\centerline{\includegraphics[width=\linewidth]{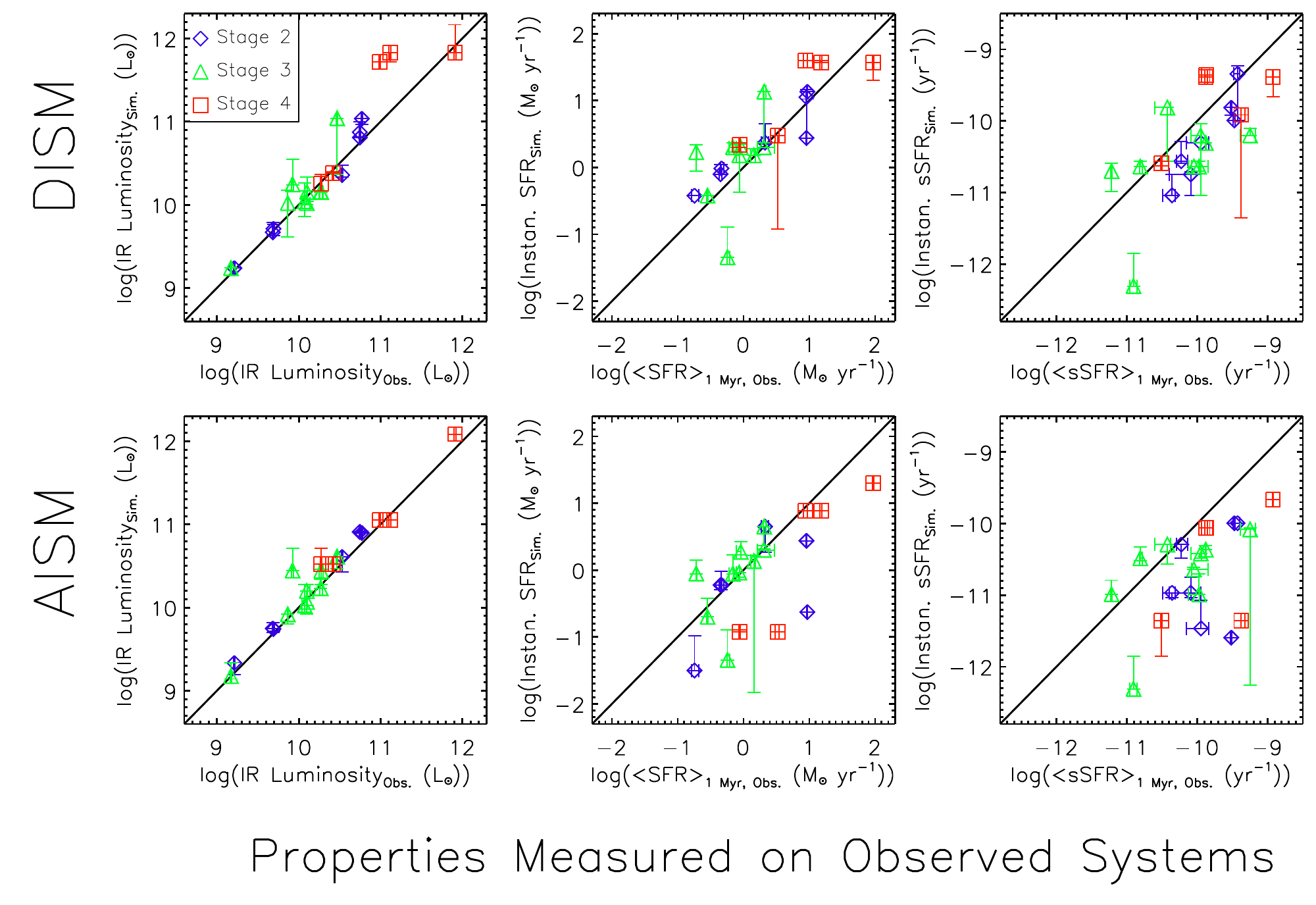}}
\caption{Comparison for the DISM (upper row) and AISM (lower row) matches of: the IR luminosity (left), SFR averaged over the previous 1\,Myr (middle), and sSFR (sSFR) derived for the observations with MAGPHYS and the property of the best-matched simulated SEDs.  The best-matched simulated SED's value is plotted and the vertical error bars show the range of the parameter for the set of best matched simulated SEDs. Blue diamonds are Stage 2 (weakly interacting) systems, green triangles are Stage 3 (moderately interacting) systems, and red squares are Stage 4 (strongly interacting) systems. We find that the parameters are typically well recovered, given the coarseness of our parameter space coverage. The DISM matches often have higher SFR; a possible reason for this effect is that under the DISM assumption, older stars are less obscured and thus contribute less significantly to the IR luminosity than when the AISM assumption is used.}
\label{par_comp1}
\end{figure*}

We compare SFR in the middle panels of Figure \ref{par_comp1}. While MAGPHYS can determine several SFR averaged over different timescales, we chose to compare the average over 1 Myr, the shortest available, as being closest to the instantaneous SFR recorded during the hydrodynamic simulation. We find fairly good agreement for both AISM and DISM matches, although the AISM matches have more systems that not as well recovered. 

Several moderately interacting systems (shown in green) also display interesting offsets. The most striking is NGC\,4647/4649 whose simulated SFR is particularly low at around log(SFR/(M$_{\odot}$ yr$^{-1}$))$=-1.3$ compared to the observed SFR of log(SFR/(M$_{\odot}$ yr$^{-1}$))$=-0.2$. NGC\,4649 is a large elliptical galaxy, which we do not have in our simulated galaxies. It was not detected at most MIR-FIR bands, therefore the system has little constraints on that region of the SED. It is therefore not surprising that its SFR is not well recovered. In contrast, the SFR associated with the best simulated matches to NGC\,3185/3190 is higher than the observed SFR (log(SFR/(M$_{\odot}$ yr$^{-1}$))$=-0.1$ vs log(SFR/(M$_{\odot}$ yr$^{-1}$))$=-0.7$). This is likely related to the significant over-estimation in the UV of the best matches (see Figure \ref{int_fits}), where the best matches were not able to find as heavily obscured a system as one containing the nearly edge-on NGC\,3190.  We find that M51's log(SFR/(M$_{\odot}$ yr$^{-1}$))$=0.3$ is better recovered by the AISM matches, as expected from Figure \ref{int_fits}. The distribution of sSFR (right panels of Figure \ref{par_comp1}) is fairly similar to the distribution of SFR, but with a weaker trend due to the degree of recovery of the stellar mass. 

\begin{figure*}
\centerline{\includegraphics[width=\linewidth]{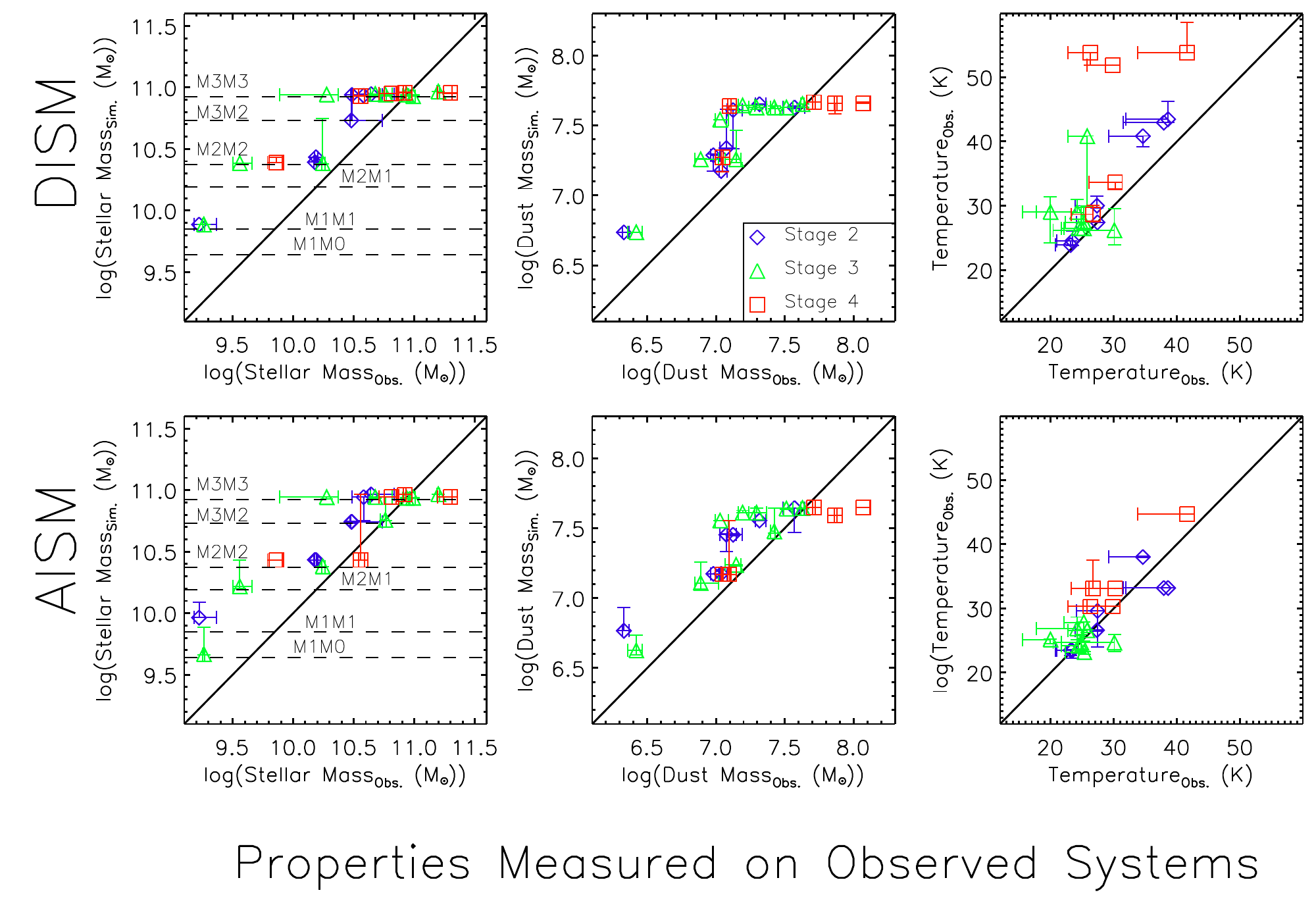}}
\caption{Comparison for the DISM (upper row) and AISM (lower row) matches of: the stellar mass (left), effective dust mass (middle), and dust temperature (right) derived for the observations with MAGPHYS and the property of the best-matched simulated SEDs.  The temperatures compared in the right panels are the temperatures of a $\beta=1.5$ optically-thin modified blackbody fit to the simulated and observed SEDs. The best-matched simulated SED's value is plotted and the vertical error bars show the range of the parameter for the best-matched simulated SEDs. The horizontal error bars are the uncertainty derived with MAGPHYS (masses) and from the choice of the beta used in the temperature fits. Blue diamonds are Stage 2 (weakly interacting) systems, green triangles are Stage 3 (moderately interacting) systems, and red squares are Stage 4 (strongly interacting) systems. The horizontal lines in the stellar mass plot show the initial stellar mass of the labeled simulations. The simulated stellar masses increase only modestly over the course of a simulation because the initial gas fractions are relatively low and no additional gas is supplied. We find that the sparsity of the parameter space coverage in mass makes it difficult to determine the degree to which the stellar and dust mass are recovered, although more massive systems tend to find matches with the SEDs of simulations of more massive interactions. We find that the AISM SEDs better reproduce the effective dust temperatures of the observed systems. }
\label{par_comp2}
\end{figure*}

\subsubsection{Stellar  Mass}

In the left panels of Figure \ref{par_comp2}, we compare the stellar masses of the observed systems to those of their best-matched simulated counterparts. More massive systems are better matched by simulations of more massive galaxies, which is not surprising given the broadly normalizing aspect of stellar mass and its importance in driving the intensity of an interaction. Because the simulated interactions do not gain material from their environment, their stellar mass evolves little over the course of a simulation, as they have a finite gas reservoir out of which to form stars.  This results in a sparse coverage of the stellar mass parameter space.  Additionally, our simulation suite does not have a large variety of interaction parameters or gas fractions, which, among other simulation characteristics, may impact the stellar mass of the best-matched SEDs.  Therefore, determining more precisely how well stellar mass is recovered and the cause of a possible tendency of the less massive systems to match more massive simulations is not feasible with the current suite of simulations.  Connected with this topic is the development and evolution of the so-called galaxy main sequence for star formation: SFR vs. M$_{*}$ \citep{hayward14}. 
 
 \subsubsection{Dust Mass}
 
The middle panels of Figure \ref{par_comp2} compare the simulated and observed dust masses. The distribution generally looks similar to stellar mass (i.e.,  systems with more dust are better matched to simulated SEDs calculated through a dustier environment). Although the simulated dust mass tends to be higher than observed at intermediate masses, it agrees within a factor of  $\approx2-3$, which is typically the level of uncertainty in the determination of dust mass.

\subsubsection{Effective Dust Temperature}

\begin{figure*}
\centerline{\includegraphics[width=0.97\linewidth]{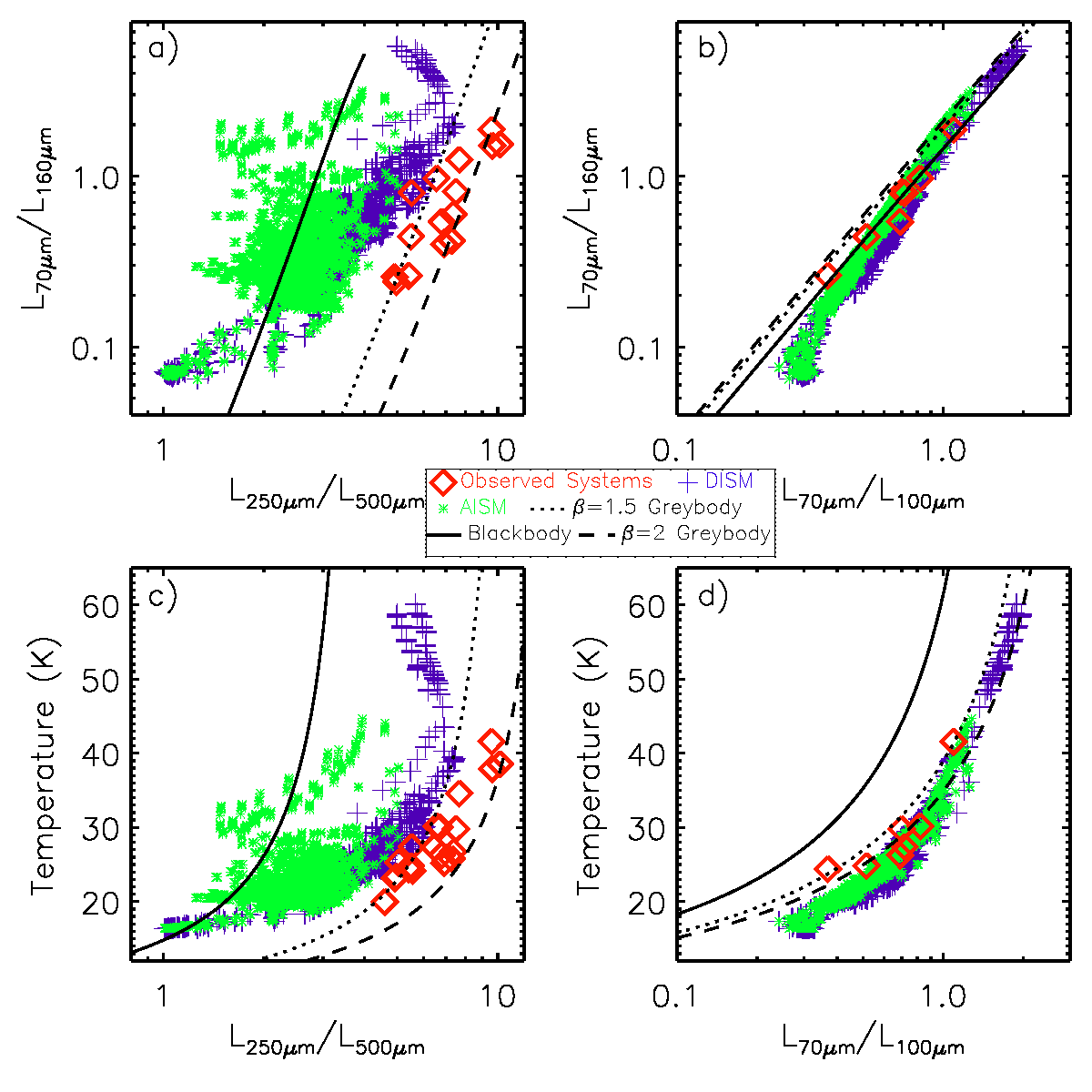}}
\caption{Comparison of the Herschel colors and derived temperatures between the observed (red diamonds), AISM (green stars), and DISM (blue crosses) datasets showing that the PACS colors (upper right) are similar, but neither simulated dataset well recovers both the SPIRE and PACS colors (upper left). In general, the simulations and the observations overlap in $L_{70}/L_{100}$ and $L_{70}/L_{160}$, but the $L_{250}/L_{500}$ ratios of the simulations tend to be less than those observed. The up-turn between 350\um~and 500\um~that can be seen in some of the AISM fits in Figure \ref{int_fits} is responsible for the low ratios of the AISM dataset. The lower panels demonstrate that the temperature, derived from a single-T $\beta=1.5$ fit to the observed and simulated photometry is much more tightly correlated with the PACS bands (right) than the SPIRE bands (left). For comparison, we include curves indicating the positions of a blackbody (solid) and two modified blackbodies ($\beta=1.5$ dotted; $\beta=2$ dashed).}
\label{colors}
\end{figure*}

\begin{figure*}
\centerline{\includegraphics[width=0.95\linewidth]{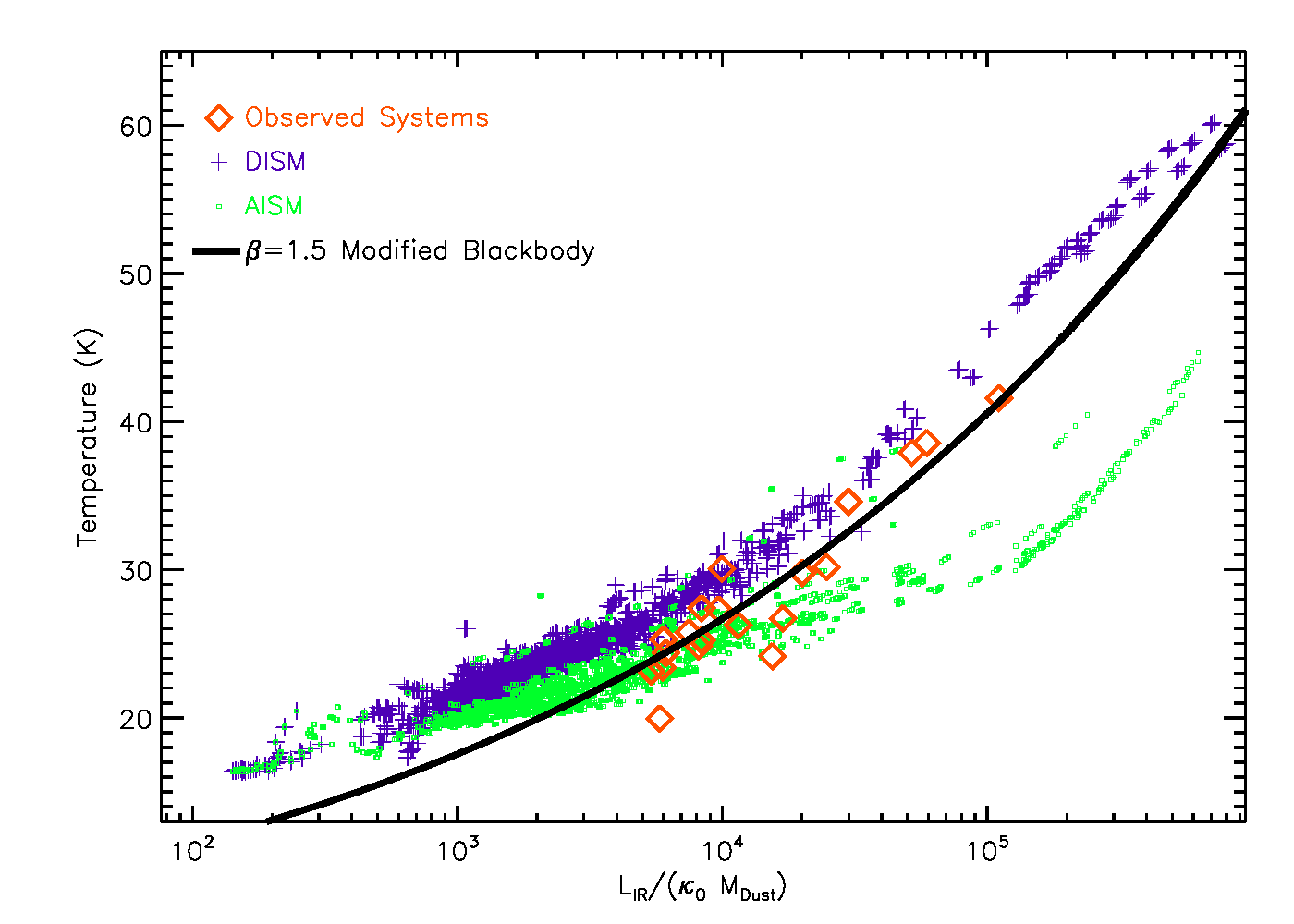}}
\caption{Effective dust temperature determined by fitting a single-temperature $\beta$=1.5 modified blackbody  versus the ratio of IR luminosity to dust mass. In red, green, and blue, we show the observed (red diamonds), AISM (green stars), and DISM (blue crosses) datasets, respectively. The observations are well modeled by an optically-thin modified blackbody (black line), as expected because the dust masses output by MAGPHYS are calculated assuming a very similar form. In the simulations, the 3-D dust distribution is a physical input, and the dust grain temperatures vary significantly depending on the local radiation field and the grain size; thus, there is no a priori reason that the effective dust temperature should behave according to the black line or, more generally, tightly depend on $L_{\rm IR}/(\kappa_0 M_{\rm Dust})$. The effect dust temperature for the DISM SEDs scale in a manner similar to that expected for a single-temperature optically-thin modified blackbody at all temperatures, while the AISM temperatures only do so at temperatures above $\sim$30\,K.  The low scatter at a given $L_{\rm IR}/(\kappa_0 M_{\rm Dust})$ value suggests that the effective dust temperature (i.e., location of the peak of the IR SED) is driven primarily by thermal equilibrium; other factors, such as the geomtry or compactness of the system, are subdominant.}
\label{tmp_lirmd}
\end{figure*}

The right panels of Figure \ref{par_comp2} compare the temperatures of $\beta=1.5$ blackbodies fit to each observed and matched simulated SEDs. We find as expected that the simulated DISM SEDs nearly always have hotter dust temperatures particularly at observed temperatures of 30-45 K. The exception is NGC\,3185/3190, whose temperature is uncertain due to the lack of PACS and MIPS FIR photometry. The AISM SEDs recover the dust temperature much better. 

Figure \ref{colors} compares the Herschel colors and the derived temperatures for the observed and simulated photometry. The simulations reproduce the observed PACS colors well but there is little overlap in the SPIRE colors, particularly between the observed and AISM photometry. This difference is primarily seen in Figure \ref{int_fits} in the up-turn in the AISM FIR photometry generally found between 350\um~and 500\um. \citet{jon10} found a similar discrepancy in the sub-millimeter colors (e.g., MIPS 160\um/SCUBA 850\um) of their simulations and the SINGS galaxies. For comparison in Figure \ref{colors}a, a blackbody ($\beta=0$) has colors that pass through the center of the AISM points (green stars); a modified blackbody of $\beta=1.5$ passes through the locus that marks the right edge of the simulation points (green and blue); and a modified blackbody of $\beta=2$ passes through the observed dataset (red diamonds). This figure suggests that the observations fit well with a model having a $\beta$ between 1.5 and 2, which for the observed SPIRE ratios corresponds to a dust single-temperature range from $\sim$15\,K to $\sim$40\,K. Both sets of ISM simulations produce SEDs whose locus of points bridges a much wider area, at low values of $L_{70\,\mu{\rm m}}/L_{160\,\mu{\rm m}}$ best represented by a $\beta=0$ (T$\sim20-30$\,K) model and at values of $L_{70\,\mu{\rm m}}/L_{160\,\mu{\rm m}}\ge 1$ best represented with a $\beta=1.5$ (T$\sim30-60$\,K) model. Similarly, Figure  \ref{colors}c, which also uses SPIRE photometry, finds the observations are best fit with a $\beta \ge 1.5$ model, while the simulations span a larger area. The observations plotted in Figure \ref{colors}a and \ref{colors}c occupy a much more restricted space because the observed SPIRE luminosity ratios is always greater than 4, while the simulated ratios fall as low as 1 (corresponding to very cold dust with T$\sim15$\,K assuming $\beta=0$). Figures \ref{colors}b and \ref{colors}d, which use only PACS photometry, show a tight correlation between the observations and the simulations, with a preference for a $\beta \ge 1.5$ model. 

In Figure \ref{tmp_lirmd}, we plot the temperature of a single $\beta$=1.5 modified blackbody fit to each simulated and observed SED against the ratio of the IR luminosity and dust mass associated with that SED. If an optically thin  $\beta$=1.5 blackbody were a perfect model for the photometry, we would expect this relation between the two quantities since (from eq. A4 of \citet{hay11}; shown as the black line in Figure \ref{tmp_lirmd})

\begin{eqnarray}
T \sim 5{\rm K} \Bigg(\frac{L_{IR}}{\kappa_{0} M_{Dust}}\Bigg)^{\frac{1}{4+\beta}}.
\end{eqnarray}
\noindent

The observed systems agree well with this model, as expected since their cool dust masses are derived by MAPGHYS assuming emission as two modified blackbodies, and in MAGPHYS, the cold dust typically dominates the total dust mass. The DISM SEDs also show temperatures that scale similarly to the blackbody model, as can be seen in the slope of its points and the tightness of its correlation. While the AISM temperatures scale with a noticeably flatter slope at low temperatures, its temperatures scale similarly above $\sim$30\,K. The tightness of these relations, for a given ISM assumption suggests that the luminosity and dust mass are the main parameters necessary in setting the FIR peak, while geometry likely only contributes to the scatter. The offset in DISM simulations may be due to the fact that the simulated dust actually has a temperature distribution rather than a single temperature, whose hot dust could drive the effective temperature a bit higher. The offset in AISM may be caused by dust self-absorption, resulting in an observed cooler temperature than the temperature of the hot core.

\subsection{Validity of Morphology-based Interaction Stages Estimates}
\begin{figure*}
\centerline{\includegraphics[width=\linewidth]{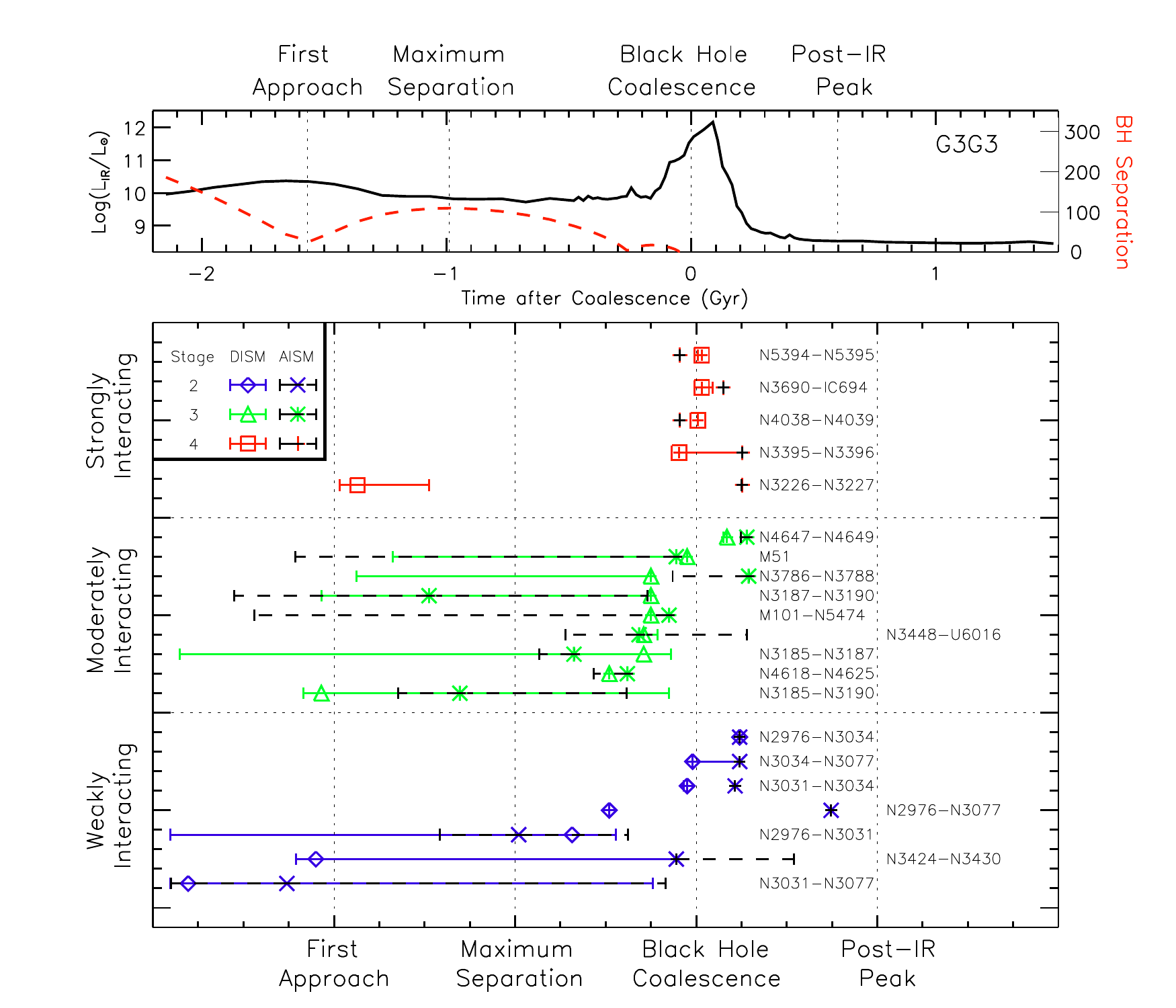}}
\caption{Comparison of the Dopita classifications with the timing of the best matches from the AISM (crosses, stars, and `x') and DISM (squares, triangles, and diamonds) simulations relative to landmarks of interaction. The top plot shows the evolution of the M3M3 simulation, showing the IR luminosity (black, solid line) and the separation of the two SMBH (red, dashed line). The vertical dotted lines identify the position of the first three landmarks identified from the black hole separation. The fourth landmark is the time at which the IR luminosity has decreased from its peak to a fairly constant value. In the bottom panel, the landmarks are equally spaced, as the time between landmarks varies for each interaction. The data points are plotted at the fractional time between two landmarks from which the best SED match originate. The ``error bars'' show the range of snapshots from which the DISM (solid color) and AISM (dashed black) SED matches originate. The vertical spread in each class is simply to aid in distinguishing the different systems.  In most cases, the matches for the galaxies classified as strongly interacting come from the near-coalescence phase of the simulations, only. For the other interaction stages addressed in this work (weakly and moderately interacting), the integrated SED alone is insufficient to identify the interaction stage.}
\label{sim_time}
\end{figure*}

\begin{figure*}[thp]
\centerline{\includegraphics[width=\linewidth]{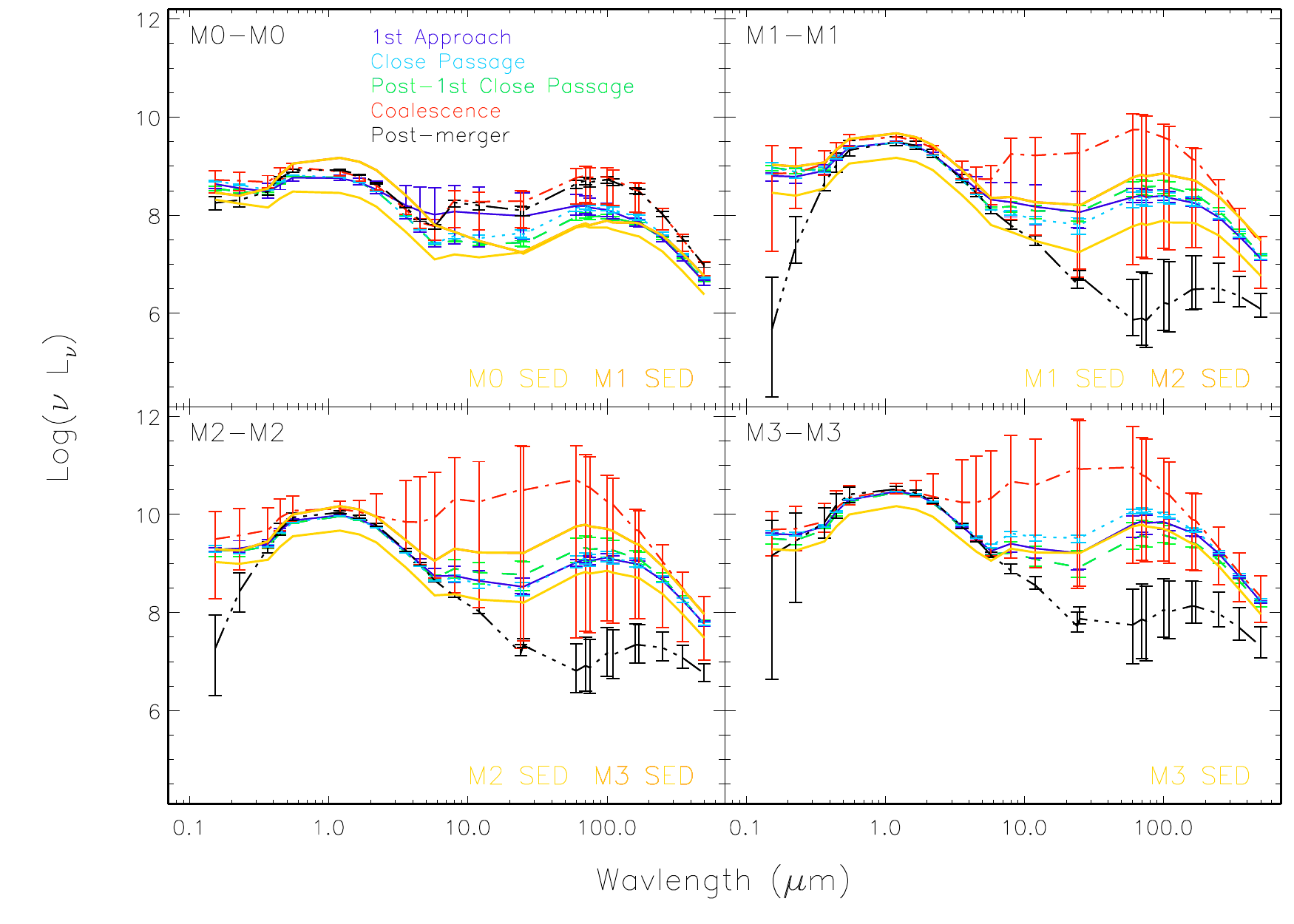}}
\caption{Median DISM SED for five stages in the equal mass mergers: the initial approach, first passage, maximum separation after the first passage, coalescence, and post-merger for one of the viewing angles. The AISM SEDs are generally very similar, albeit with enhanced MIR emission during coalescence, particularly in the G2G2 and G3G3 simulations. The``error" bars show the range over each stage. Note that the SEDs vary most during the coalescence phases because of the rapid variation in the SFR, IR luminosity, and dust mass during this phase. As noted above, the times of the strongest interactions tend to feature district SEDs otherwise these SEDs show that even in major mergers, which exhibit the greatest activity, the SED alone is generally insufficient to identify the interaction stage, except during the most active phases and in the passively-evolving post-merger stage; the latter typically have significantly lower UV and FIR luminosities because star formation has been quenched.}
\label{med_seds}
\end{figure*}

Most interacting systems are characterized by their morphological distortion. The \citet{dopita02} classification scheme uses this method. Weakly interacting (Stage 2) systems are only mildly distorted, while strongly interacting (Stage 4) systems are significantly distorted. In Figure \ref{sim_time} we examine how well this classification works as a proxy for the location on the interaction sequence. However, each simulation proceeds at a unique pace governed by its particular combination of galaxy masses, mass ratios, and initial separation. Hence, just comparing time relative to a single fixed event (such as coalescence) does not necessarily provide a useful comparison. For example, 1\,Gyr before coalescence, M3M3 is approximately at the maximum separation after first passage, but M2M2 has not yet had its first pericenter passage. 

Many interactions share a number of common landmarks along the interaction sequence. We define four signposts to examine the relative location of the best matches. The vertical lines in the top panel of Figure \ref{sim_time} indicate the times of our landmarks in the M3M3 simulation. The first three are defined based on the separation of the central SMBHs, which acts as a proxy for the separation of the galaxies. The first two landmarks are the first close approach and the moment of maximum separation after that initial passage. The simulations have a variety in the number of close approaches, which tend to increase with stellar mass ratio. The third landmark is the moment at which the two SMBHs coalesce. However, in the major mergers where the increase in IR luminosity is pronounced, the peak luminosity occurs after that coalescence \citep[c.f.][]{hop06}.  Therefore, we define a fourth landmark when the IR luminosity has decreased from its peak to a low but steady level.	

We determine where each set of best matches fall between the landmarks. In the bottom panel of Figure \ref{sim_time}, we plot the range covered by these matches from the DISM and AISM comparisons against the Dopita system classification. We generally do find that the strongly interacting (strongly distorted) systems originate from the period of black hole coalescence and peak IR emission. The exception is NGC\,3226/3227's DISM matches. We see a similar separation between the AISM and DISM matches in NGC\,2976/3077. Neither system has UV photometry and the shape of the UV simulated emission differs between the AISM and DISM matches. 

For the other ``Dopita classes'', we do not find a clear trend of evolution along the interaction sequence with the morphologically determined classes, demonstrating that the SED alone is insufficient to uniquely determine the interaction stage. The moderately interacting systems span most of the sequence, as do the weakly interacting systems, several of which cluster mostly in the same period as the strongly interacting systems.  There is, however, a strong caveat: our observational sample is fairly small. Although we have seven pairs of weakly interacting galaxies, six originate from the same system (the NGC\,3031-NGC\,3034 system), and NGC\,3034 is by no means a typical galaxy in the early interaction stages. Similarly, our simulations make a good beginning at spanning the properties of our observed samples, but we would not claim that we are simulating counterparts specific to any of our observed system or indeed sampling the possible interaction types with great resolution. Further, there is the additional complication that interaction stages defined by degree of morphological distortion suffer from some degeneracy. Previous simulations of interacting systems have also demonstrated that the appearance at a given time during an interaction depends not only on the specific geometry of the encounter, but also on the masses, metallicities, gas contents, and previous interactions of the progenitor galaxies \citep[e.g.,][]{dim07}.

\subsection{Evolution of SEDs in Major Mergers}

We use the landmarks described in Section 5.2 to define five interaction stages: the initial approach, the first close passage, the separation post-1st passage, coalescence, and the relaxation period after the merger. For each stage, we determine the median SED for each of the equal-mass mergers (e.g., M0M0) as seen from one of the viewing angles. There is little difference between the median SEDs for different viewing angles because the galaxies do not share an equatorial plane, so there is no preferred viewing angle even early in the interaction. The UV has the greatest variation with viewing angle due to the obscuring effect of dust at those wavelengths; the range covered by the camera angles is typically $\sim$0.2-0.5\,dex in $\lambda$~L$_{\lambda}$ increasing to almost an order of magnitude for about 50\,Myr $\sim$200\,Myr after coalescence in M2M2 and M3M3. 

In Figure \ref{med_seds}, we show the median SEDs for the four equal-mass mergers.  We find that the coalescence stage typically has more luminous IR emission than the other stages, although it is also the stage with the highest variation in IR emission.  Although the median SEDs are broadly similar in shape, there are a few notable differences. We find a slight enhancement in the MIR-FIR after the 1st close passage in the two intermediate mass mergers M1M1 and M2M2. Similarly, the post-merger stage, during which the system has become an elliptical, show noticeably lower  UV and FIR in the three massive mergers. For comparison, we include SEDs of the isolated simulations with similar stellar masses to each panel of Figure \ref{med_seds}. We find the median SEDs of first three interaction stages are typically bracketed over the observed wavelength range by the SEDs of isolated galaxies with lower and higher stellar masses than the total mass in the interaction, suggesting that these stages may not be distinguishable from isolated galaxies purely from the SED.

We conclude that SEDs can only be used to identify systems in the most active phases of interaction or that have reached the post-merger stage in the more massive interactions. Figure \ref{med_seds} shows the large range of variation in the SEDs of the coalescence stage. As a result, a system in the earliest interactions stages can have an SED very similar to a system in the less active portions of coalescence. Therefore, systems in these early stages can not be identified purely from their SEDs. However, systems in the most active phases of coalescence in major mergers, during which the MIR-FIR emission increases greatly relative to the NIR emission can be identified. Similarly, systems that have become a post-merger elliptical and hence have very little FIR relative to their NIR emission can also be identified.

\section{CONCLUSIONS}

We presented the first systematic comparison of SEDs of observed and simulated interacting galaxies. Our sample of 31 galaxies was observed in up to 25 bands including \emph{GALEX}, \emph{Spitzer}, and \emph{Herschel}. We created a suite of \gadgetthree hydrodynamic simulation of four galaxies evolving in isolation and the ten pair-interactions evolved from first passage through coalescence to the post-merger stage. Simulated SEDs were calculated using \sunrise for two different treatments of the ISM's multiple phases, and sets of the best-matching of these SEDs were determined for each observed pair. From our comparison of the simulated and observed SEDs, we conclude that:\\

(1)~~For most observed systems, at least one mock SED from the simulations provides a reasonably good (often statistically acceptable) fit to the observed SED. The best-matched SEDs generally come from the same simulation, often the major mergers. They tend to cluster around coalescence and the constraint in timing is tightest for the most evolved systems. We did not find, however, that best matches preferentially come from a certain viewing angle.

(2)~~Neither treatment of the sub-resolution ISM is preferred for our sample as a whole. Some interactions are better matched when only the diffuse dust was used in the radiative transfer calculations (DISM) while others have SEDs which are better reproduced especially in the FIR when the total dust mass is used in the calculations (AISM). Half of the observed interactions are equally well matched by the two sets of simulated SEDs.

(3)~~The best matches recover IR luminosity and SFR fairly well. 

(4)~~Stellar and dust masses show indications that more massive (or dustier) systems tended to be matched by simulations of more massive (or dustier) galaxies, but greater coverage of the simulated parameter space is necessary to reliability recover stellar and dust masses. 

(5)~~The DISM SEDs have dust that is typically hotter than the observed systems. The temperature is better recovered with the AISM matches. 

(6)~~Our SED matching techniques is able to reliably identify the interaction stage of the strongly interacting systems near coalescence. In contrast, the less strongly interacting morphology-based classes cover a wide range of interaction stages in their best matches. This suggests that the integrated SED alone is typically insufficient to identify the interaction stage, except for the most strongly interacting systems and the passively evolving merger remnants. 

(7)~~The SEDs of the simulated systems in the different stages are generally quite similar, supporting our previous conclusion. The two exceptions are first, the passively evolving merger remnants (which exhibit markedly less UV and FIR emission because of quenching), and second, the most strongly interacting systems. For the latter, the SED varies significantly because of the rapid variation in the SFR and AGN luminosity. \\

Our study improved upon the comparison of \citet{jon10} who tested  \sunrise against the SEDs of the SINGS galaxies by addressing two of the five issues of concern they raised: the small range of simulations available to them to fit normal galaxies, and the lack of treatment for the cold ISM. Regarding their first issue, we found in our larger range of systems that the simulations and \sunrise do a good (albeit not perfect) job of fitting observed SEDs over a much larger range than observed with SINGS. Additionally, our study focused on the more complex conditions in interactions while \citet{jon10} examined how well isolated galaxies were reproduced. We concur with their assessment that this simulation process yields realistic results, although there remain areas of possible improvement such as a more realistic treatments of processes like star formation at smaller scales. Regarding the issue of cold dust, their Figure 16, for example, showed clearly that the observed MIPS 70\um~and 160\um~data points lie well away from the simulations colors, indicating the incomplete treatment of cold dust in the simulation. Our PACS and SPIRE data similarly demonstrates (e.g. Figure \ref{int_fits}) that very often the FIR-submillimeter points are poorly reproduced, although our fits are much closer than the ones in the \citet{jon10} paper. We compared the dust using two treatments of sub-resolution ISM structure in the simulations; we found that some galaxies match better with one treatment, while others are better matched with the other. With the current simulations, the sub-resolution structure of the ISM remains a significant modeling uncertainty (see Figures \ref{sunrise} and \ref{stdev_sunrise} and the associated discussion). However, in future work, this uncertainty can be reduced considerably through the use of higher-resolution merger simulations that include a more-realistic multiphase ISM \citep[e.g.,][]{hop13a, hop13b}. 

This study has demonstrated that with even a relatively modest number of simulations, it is possible to match the SEDs of observed interacting galaxies with one or more mock SEDs from the simulations. Furthermore, the best matches tend to come from relatively constrained regions of the physical parameter space, which suggests that the fits are reasonably non-degenerate. While a more expansive library of simulations may yield better matches to some of our observed systems, these comparisons, together with \citet{jon10},  demonstrate the feasibility of directly inferring physical quantities of galaxies by direct comparison with forward-modeled mock SEDs rather than through traditional SED modeling, assuming that the simulations span a sufficiently large parameter space with sufficiently fine sampling. Although creating the template SEDs using this forward-modeling approach is orders-of-magnitude more computationally expensive than standard SED modeling, the advantage is that this approach is predictive in the sense that the SEDs are calculated self-consistently from hydrodynamical simulations; the predictive nature of the forward-modeling approach yields additional value beyond inferring physical parameters of the galaxies, such as the possibility of examining the potential evolution and past of these systems.

This work has utilized only the integrated SEDs of the simulated galaxies, but these represent only a tiny fraction of the information available from the simulations. In future work, we will more fully utilize the wealth of data available from the simulations by comparing the simulated and observed galaxy morphologies, for example, and determining what morphological information is encoded in the integrated SEDs.

\acknowledgements
The simulations in this paper were performed on the Odyssey cluster supported by the FAS Research Computing Group at Harvard University. LL and HAS acknowledge partial support from NASA grant NNX12AI55G and JPL RSA contracts 717437 and 717353. CCH is grateful to the Klaus Tschira Foundation for financial support and acknowledges the hospitality of the Aspen Center for Physics, which is supported by the National Science Foundation Grant No. PHY-1066293. This work was based on archival data obtained from the Spitzer Science Archive, the Mikulski Archive for Space Telescopes (MAST), the Swift data archive, and the Herschel Science Archive. \emph{Herschel} is an ESA space observatory with science instruments provided by European-led Principal Investigator consortia and with important participation from NASA. \emph{Spitzer} is operated by the Jet Propulsion Laboratory, California Institute of Technology under a contract with NASA. \emph{GALEX} is operated for NASA by the California Institute of Technology under NASA contract NAS5-98034. This research has made use of the NASA/IPAC Extragalactic Database (NED), which is operated by the Jet Propulsion Laboratory, California Institute of Technology, under contract with the National Aeronautics and Space Administration.

\clearpage

\clearpage

\clearpage

\clearpage

\end{document}